\documentclass[sigconf,nonacm]{acmart}

\AtBeginDocument{%
  }

\setcopyright{acmlicensed}
\copyrightyear{2027}
\acmYear{2027}
\acmDOI{XXXXXXX.XXXXXXX}
\usepackage{fontawesome5}
\acmConference[Conference acronym 'XX]{Make sure to enter the correct
  conference title from your rights confirmation email}{June 03--05,
  2018}{Woodstock, NY}

\acmISBN{978-1-4503-XXXX-X/2018/06}
\author{
   Evan Pačes-Wiles\textsuperscript{1*}  
   Bryen Alvarez\textsuperscript{1*} 
   Sara Gulabani\textsuperscript{2*} 
   Akshara Kolipaka\textsuperscript{1*} \\ 
   Snehalkumar `Neil' S. Gaikwad \textsuperscript{1,2*}\textsuperscript{\faEnvelope[regular]}
}
\affiliation{\vspace{0.5em}
  \institution{\textsuperscript{*}Society-Centered AI Lab}
  \institution{\textsuperscript{1}Department of Computer Science}
  \institution{\textsuperscript{2}School of Data and Information Sciences}
  \city{University of North Carolina, Chapel Hill}
  \country{USA}
}
\email{gaikwad@cs.unc.edu}

\usepackage{xurl}
\usepackage{placeins}
\usepackage[skins]{tcolorbox}
\tcbuselibrary{breakable}
\definecolor{promptbg}{RGB}{247, 247, 248}
\definecolor{promptborder}{RGB}{229, 231, 235}
\definecolor{promptcaption}{RGB}{107, 114, 128}

\newtcolorbox{promptbox}[1][]{
    colback=promptbg,
    colframe=promptborder,
    arc=8pt,              
    boxrule=1pt,          
    left=15pt, right=15pt,
    top=15pt, bottom=15pt,
    breakable,            
    enhanced,
    fonttitle=\bfseries\color{promptcaption},
    title=AI Prompt,      
    attach title to upper,
    after title={\par\medskip\hrule\medskip},
    #1
}

\begin{document}

\title{EJAI: Judgment-Preserving Civic AI for \\ Data Center Environmental Justice}

\begin{abstract}
 Communities in the United States and around the world are contesting data centers, the physical infrastructure of AI. These disputes pose environmental justice issues: residents must weigh unevenly distributed burdens using evidence scattered across metrics, corporate claims, policy documents, and public testimony. A formative study with 10 users showed that dashboards and single-agent AI can settle questions before people examine that evidence. Here we present \textsc{EJAI}, a source-grounded system that links an interactive map, retrieved evidence, missing-data statements, and two stakeholder personas that allow people to inspect evidence and form their own policy recommendations. In a small functional benchmark of 21 queries, retrieved evidence supported 98.4\% of factual claims and the system acknowledged missing evidence in 32 of 33 cases. Persona alignment carried a grounding cost: the Corporate Executive omitted some evidence that conflicted with its assigned priorities. In a within-subjects study, 15 participants acting as residents prepared for a simulated town hall. Some checked cited AI claims while treating map metrics as official, some made conditional recommendations with safeguards such as water-use disclosure, and some found the personas simplified or biased. Our results suggest that civic AI can support preparation for public reasoning when it makes evidence, gaps, and disagreement open to inspection, and that corpus imbalance and two-sided personas remain open justice concerns.
\end{abstract}



\begin{teaserfigure}
\centering
\includegraphics[width=0.825\textwidth]{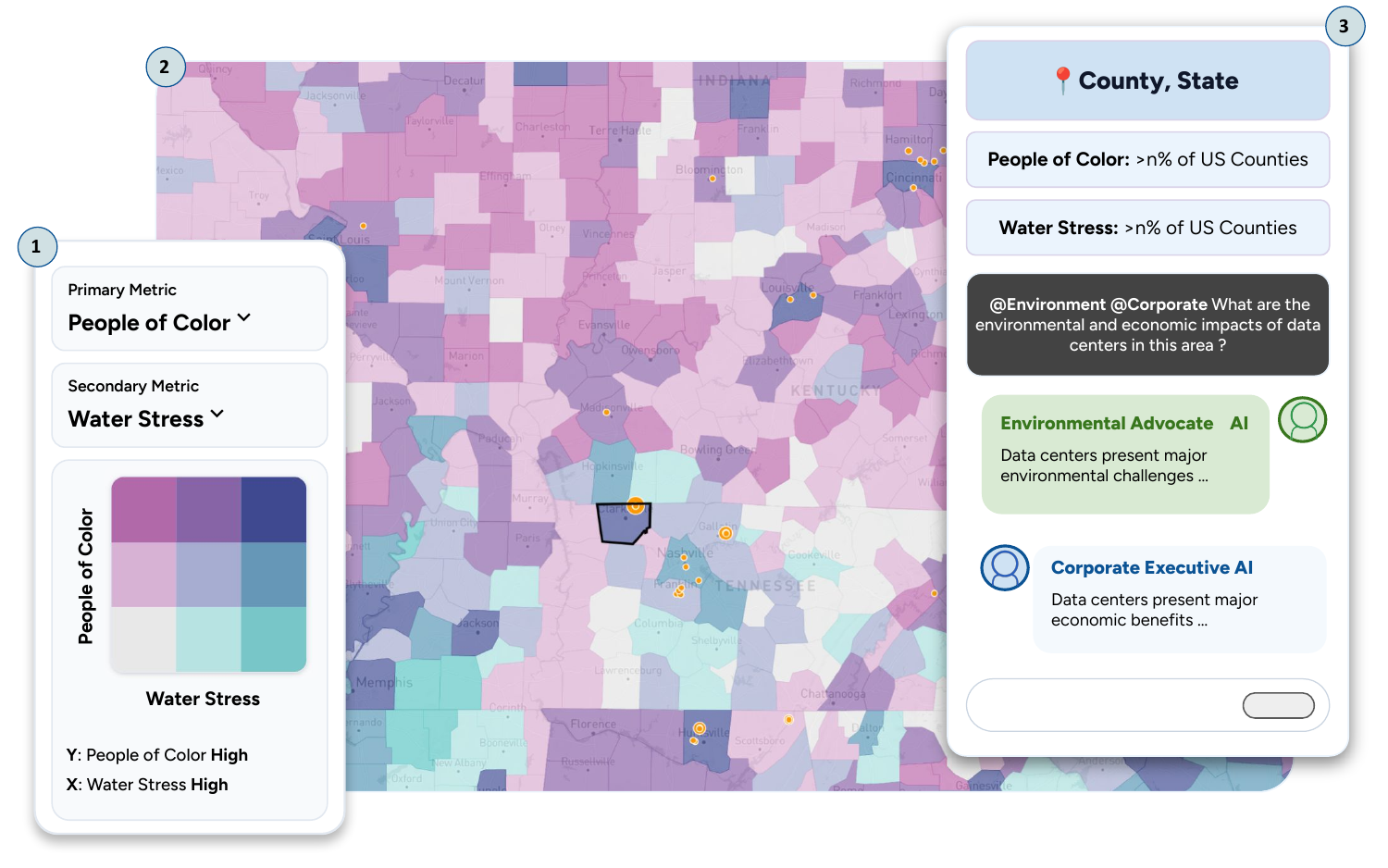}
  \caption{\textsc{EJAI} is a source-grounded multi-agent visualization system for data center environmental justice. Users can filter data centers and visualize  metrics (1) on a map of data centers (2). Clicking a county surfaces an overview of selected and outlier metrics, tracked data centers, and local quotes. Users can query the system (3),  including @-mention personas, for source-grounded responses.}
  \Description{A teaser figure of the major components of the interface. Includes a bivariate chloropleth map displaying county-level metrics for people of color and water stress, with a metric legend on the left side of the figure. A county is selected, showing high levels of both. On the right side of the figure, a chat window is shown grounded in the selected county, with a brief illustration of a conversation between a user and two personas: an Environmental Advocate and Corporate Executive.}
  \label{fig:teaser}
\end{teaserfigure}

\maketitle

\section{Introduction}
As AI expansion drives demand for data center infrastructure \cite{li2025making}, several North Carolina governments have paused approvals for data centers. On October 6, 2026, the Raleigh City Council voted 7--1 to pause approvals for certain large data centers for six months, as residents asked how working-class neighborhoods would be protected and how they would have a real voice \cite{wuncRaleigh2026}. In 2026, Durham County adopted a nine-month pause \cite{wuncDurham2026}, Orange County adopted a one-year pause  \cite{wuncOrange2026}, and Charlotte extended its pause to as late as October 2027 \cite{bisnowCharlotte2026}. Developers and some local officials emphasize tax revenue, jobs, and investment, while residents and advocates raise concerns about water use, electricity demand, land conversion, emissions, noise, and utility costs \cite{ngata2025cloud, george2026ecological}. These disputes raise environmental justice concerns because burdens and benefits may fall unevenly across communities \cite{ngata2025cloud, george2026ecological, hajiesmaili2025toward}. We treat data center siting as an environmental justice problem grounded in broader concerns with distribution, recognition, and participation. Residents must evaluate burdens using evidence that is fragmented, technical, incomplete, and shaped by institutional interests.

Reasoning about these disputes requires connecting metrics, facility records, corporate claims, policy documents, and community testimony \cite{fusi2023unveiling, ngata2025cloud}, work often performed by advocates, journalists, and community organizations as civic data intermediaries \cite{qiao2025near, davis2021participatory}. Dashboards can show where facilities are located and which counties have high burdens \cite{10.1145/3558099, mcintyre2025extreme}, but they typically present metrics without local context. LLM assistants can explain terminology and summarize documents \cite{liuSeleniteScaffoldingOnline2024}, but authoritative answers may encourage overreliance without verification. Neither is designed to help residents examine evidence while forming their own view. Could a resident who learns of a proposed data center at a public meeting get help inspecting the evidence for their county, seeing what is missing, comparing how stakeholders interpret the same evidence, and reaching their own recommendation?

Critical research on AI's societal impacts has documented the environmental costs of AI infrastructure \cite{data_ecofeminism, hernandez2024mapping, li2025making} and argued that the public should have democratic control over how AI is developed \cite{bogiatzis2024democraticcontrol, data_ecofeminism}. Within HCI, scholars have called for balancing efficiency and ease of use with environmental justice \cite{soden2025climate}. We build on this work by asking how an interactive visualization--LLM system can help the public scrutinize the infrastructure AI depends on and support civic judgment about it. We call this design aim \emph{judgment-preserving civic AI}: LLM systems that help people inspect evidence, recognize uncertainty, compare interpretations, and form their own civic recommendations. Drawing on Sen's capability approach and his account of democracy as public reasoning \cite{sen1999development, sen2009idea}, we design for four civic capabilities: engaging with local evidence; understanding what it means and what it leaves out; interacting with competing interpretations; and preparing to act through questions, conditions, and recommendations.

Dashboards and LLM assistants carry different risks for this aim. Dashboards can leave users to infer what metrics mean, how they relate to a proposed facility, and what evidence is missing; practitioners report that frontline residents can find existing environmental justice screening tools confusing, or alarming in how they represent their communities, rather than empowering \cite{dougan2026environmental}. Single-agent chat can detach interpretation from local context \cite{qiao2025near, CivicDataSeams2026}, obscure missing evidence \cite{kim2025fostering, warren2026evidence}, and make contested decisions appear settled \cite{schmude2026contestability, felinAIActorspecificDecisions2026a}. In a formative study with 10 users, participants with a map and documents filled evidence gaps with assumptions, while those who also had a single-agent assistant generally used it as an answer source, with little source verification. Both patterns risk premature closure, in which people settle on an interpretation before examining the evidence behind it.

To address these risks, we frame AI as a data intermediary that handles bounded evidence work, such as defining metrics, summarizing sources, and surfacing gaps. This framing raises three challenges. Computationally, a system must combine metrics, facility records, documents, and testimony, state what those sources cannot answer, and keep stakeholder personas distinct while grounded in shared evidence. Interactionally, it must reduce effort while leaving interpretation with the user. Normatively, it must account for sources that carry different interests and authority, and for personas that can flatten the people they represent.

We thus present \textsc{EJAI}, a source-grounded multi-agent visualization--LLM system for environmental justice sensemaking around data center development. Its technical contribution is a coordination pipeline that binds map interaction to evidence retrieval and generation. Map selections pass to a retrieval planner, responses carry inline citations, missing-data statements, and map-layer suggestions. Two stakeholder personas, the Environmental Advocate and the Corporate Executive, debate the same evidence in turn before a deliberation matrix summarizes the exchange.

\begin{figure*}[!ht]
    \centering
    \includegraphics[width=\linewidth]{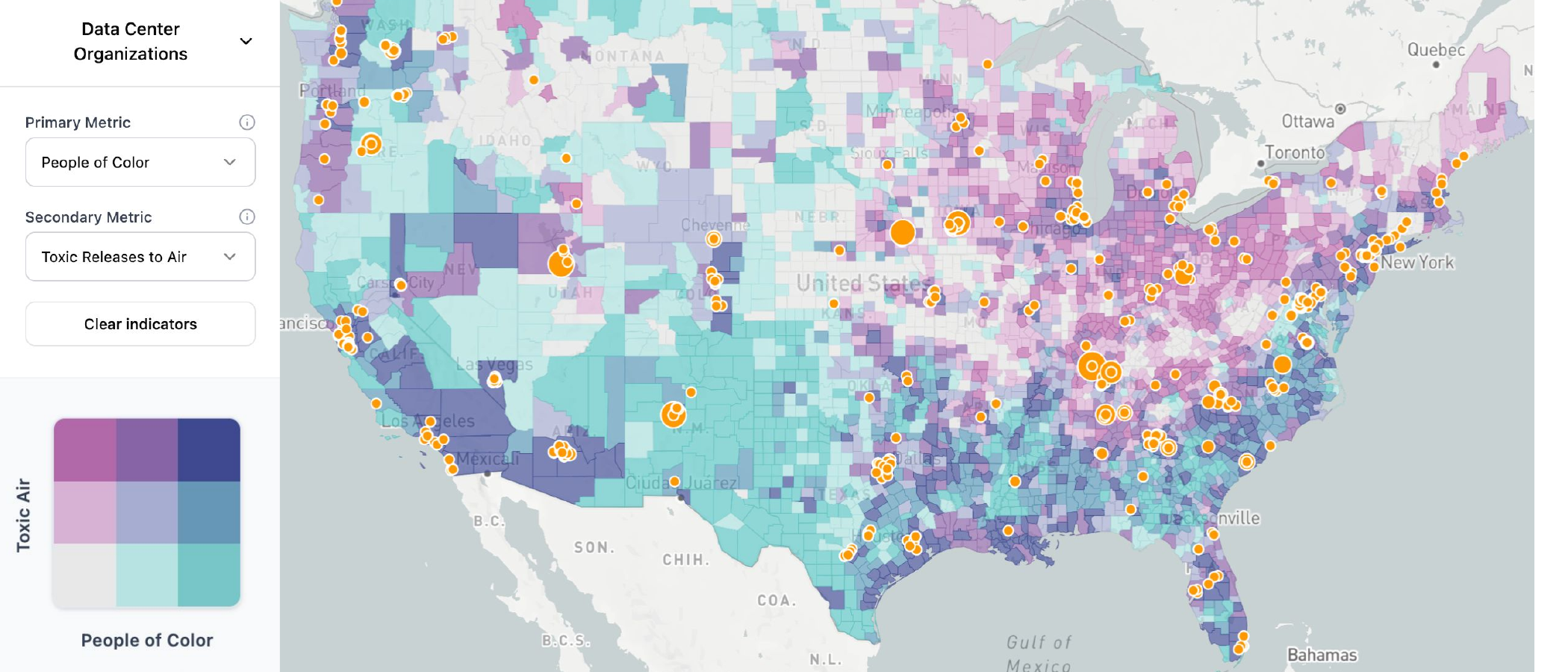}
    \caption{\textsc{EJAI} map. Environmental and socioeconomic indicators are visualized at the county level. Users can combine two metrics in a bivariate view and inspect tracked data center facilities across the United States.}
    \label{fig:map}
    \Description{A map of the United States displaying county-level data using a bivariate choropleth color scale. Points representing tracked data center facilities are scattered across the map, with clusters in some regions. Selected metrics and a legend are shown beside the map.}
\end{figure*}

We evaluate \textsc{EJAI} through a technical benchmark and a within-subjects user study. In a small functional benchmark of 21 queries, retrieved evidence supported 98.4\% (126/128) of factual claims (90.6\% under a stricter judge), a source cited in the same sentence supported 89.1\% (114/128), and the system acknowledged missing evidence in 32 of 33 cases. The user study compared a map-only dashboard, a single-agent retrieval-augmented chatbot, and \textsc{EJAI} with 15 participants acting as residents in a simulated town hall preparation task. Many made conditional recommendations with safeguards such as water-use disclosure and community benefit agreements, and questioned whose voices the testimony represented.

This paper makes four contributions. First, it identifies two forms of premature closure in environmental justice sensemaking: assumption-driven closure with dashboards and authority-driven closure with single-agent AI. Second, it introduces judgment-preserving civic AI, a design approach in which LLM systems handle evidence work and leave civic interpretation with users. Third, it presents \textsc{EJAI}, whose coordination pipeline connects map state to evidence retrieval, cited generation, missing-data statements, and persona deliberation. Fourth, it shows that aligning personas with assigned stakeholder priorities carries a grounding cost, with 85.2\% sub-question coverage for the Corporate Executive and 96.3--100\% for two-persona exchanges. Additionally, it reports how some participants checked AI claims, trusted map metrics, rehearsed disagreement, and made conditional recommendations. Together, these contributions move from documenting the costs of AI infrastructure toward tools for examining them.

\section{Related Work}
 
In this section, we connect three threads of research: civic data and environmental justice, visualization--LLM interaction, and multi-agent deliberation.
 
\subsection{Civic Data and Environmental Justice}
 
Civic data research helps residents recognize local problems and take part in public decisions. CitiSense and Smell Pittsburgh support community air quality monitoring \cite{balesCitiSenseMobileAir2012, hsuSmellPittsburgh2020}, and Regulation Room helped users comment on proposed U.S. federal rules \cite{farina2011rulemaking}. Since public datasets require domain knowledge and technical effort \cite{young2022building, DisempoweredbyData}, civic data intermediaries translate them into local concerns \cite{Puussaar2018OpenDataCivicAdvocacy, Erete2016StorytellingData}, and recent work distinguishes institutional ``far'' data from locally situated ``near'' data \cite{qiao2025near}. Environmental justice scholarship shows that burdens track race and class \cite{bullard1990dumping}, extends justice from distribution to recognition and participation \cite{schlosberg2007defining}, and treats whose knowledge counts as a question of justice \cite{fricker2007epistemic}. Although people rely on environmental justice screening tools to contest facility siting and draft public comments, these tools often privilege opaque abstract metrics over local and plural forms of knowledge \cite{dougan2026environmental}. Therefore, recent research calls for future environmental justice tools to operationalize trust through transparent provenance, empowerment through qualitative local knowledge, and accountability by bridging data with action \cite{dougan2026environmental}. Furthermore, the Toronto Water Atlas demonstrates that environmental platforms must sustain multiple ways of knowing \cite{agrawaalTorontoWaterAtlas2026}. The capability approach evaluates such systems by what people are able to do with them, and accounts of public reasoning place open discussion of evidence and values at the center of justice \cite{sen1999development, sen2009idea}.

Recent work asks how communities can take part in decisions about AI's environmental costs \cite{data_ecofeminism, bogiatzis2024democraticcontrol, hernandez2024mapping}. To date, this work has largely centered on data that communities can see or collect. Data center disputes add a further requirement: site-level facts such as water
use \cite{li2025making} and operational details \cite{young2026clouds} are often unavailable, so gaps in the record are part of the dispute. Practitioners
similarly report that outdated or undisclosed data layers erode their trust in screening tools \cite{dougan2026environmental}. Each justice dimension is at stake in who bears burdens, who gets a voice, and whose testimony counts. \textsc{EJAI} therefore places testimony beside county metrics, labels source types, and states when evidence cannot answer a question.
 
\subsection{Visualization--LLM Interaction}

Visualization research shows that framing shapes what people conclude \cite{dorkCriticalInfoVisExploring2013, Kong2018FramesSlants}, and that the way missing data is shown affects perceived data quality and confidence \cite{song2019wheres}. LLMs extend this work: LIDA generates visualizations from data summaries and goals \cite{dibia2023lida}, and Data Formulator lets authors define data concepts in natural language or by example \cite{wang2023dataformulator}. Retrieval-augmented generation pairs models with retrieved passages \cite{lewis2020retrieval}, and Ragas proposes measures of its faithfulness and relevance \cite{es2023ragas}. These systems also carry known risks. LLMs can spread misinformation and reinforce existing biases \cite{dangers_of_language_models}, people may over-rely on AI outputs \cite{10.1145/3449287, overreliance_on_ML}, and audits have found unsupported claims in cited responses \cite{liu2023evaluating, magesh2025hallucination}.

These tools were largely designed for analysts, authors, and information seekers with a defined task. Civic decisions add a further requirement: the same indicator can carry different meanings for residents, officials, and advocates \cite{felinAIActorspecificDecisions2026a, qiao2025near}, and a missing fact can matter as much as a retrieved one. \textsc{EJAI} extends visualization--LLM interaction to this setting by treating map state and chat as shared context. Our benchmark extends RAG evaluation by scoring citation accuracy separately from faithfulness and by withholding evidence types to measure acknowledgment of gaps.
 
\subsection{Multi-Agent Deliberation}
 
Multi-agent LLM systems assign roles to divide work and produce alternative interpretations \cite{wu2023autogenenablingnextgenllm, li2023camelcommunicativeagentsmind}. Perspectra and ChoiceMates give users control over which agents they consult \cite{liuPerspectraChoosingYour2025, parkChoiceMatesSupportingUnfamiliar2026}, and the Habermas Machine helped more than 5,000 UK participants find common ground \cite{tessler2024ai}. Related research asks how closely conditioned models reflect human groups \cite{argyle2023out} and shows that personas can flatten groups and reflect stereotypes \cite{ai_imitation, default_personas}.
 
This work has largely aimed at task performance, ideation, or consensus. Infrastructure disputes add a different requirement: disagreement over values is part of the record, so a system must keep that disagreement visible while tying each position to shared evidence. \textsc{EJAI} therefore gives two personas the same evidence, lets the second respond to the first, and summarizes the exchange in a deliberation matrix. Our benchmark adds Non-Collapse and Rebuttal Grounding to measure these exchanges.
 
We argue that civic AI should support public reasoning by making evidence, gaps, and disagreement open to inspection while keeping judgment with the public.

\section{Formative Study and Design Goals}

To learn what this support should do and understand community design needs, we deployed an exploratory system to study how users reasoned across spatial indicators, policy documents, corporate claims, and AI explanations. We asked: \textbf{How does AI-mediated guidance across spatial environmental data and policy or corporate documents shape environmental justice reasoning compared with access through a dashboard and PDF documents?}

\subsection{Method}

We recruited $N=10$ participants through word of mouth and email. Participants had no formal expertise with EJScreen \cite{EJSCREENEnvironmentalJustice} or environmental justice analysis and were screened for basic data literacy. In a between-subjects design, five used a spatial visualization and document corpus, and five used the same materials plus a single-agent AI assistant that could answer questions, explain metrics, summarize evidence, and prompt reflection.

Participants completed three think-aloud tasks: find information with ground-truth answers, evaluate corporate environmental claims against public indicators and reports, and make a recommendation about a hypothetical data center expansion. Two researchers independently coded the transcripts and resolved disagreements to develop a shared codebook, and we coded treatment chat logs by query type. Given the exploratory design and small sample, we report qualitative observations.

\subsection{Findings}

\subsubsection{AI supported information seeking more than critical reflection}

Participants with the assistant used it to clarify terminology, interpret unfamiliar indicators, and locate information. P10 described how the AI corrected their interpretation: "[AI] helped me realize that what I thought baseline water stress was was not actually what [I thought] it was." Participants rarely used it for Socratic reflection. Conversation logs showed a common pattern: participants asked a factual question, received an explanation, and moved to another query, with little source verification or comparison of interpretations. P9 reported consulting documents less often: "[The assistant] made me look through that a lot less, because I feel like the AI is just very tempting." We call this pattern \emph{authority-driven closure}: the assistant's answer stood in for checking how that answer was constructed. P5 was an exception and used the assistant's suggestions to find indicators to compare against the visualization.

\subsubsection{Incomplete spatial evidence encouraged assumption-driven inference}

When the available evidence did not answer their questions, participants substituted other indicators or relied on assumptions. When P2 could not find freshwater withdrawal data, they used another water indicator while recognizing the uncertainty: "I don't see [freshwater withdrawals] here, so I'm just gonna look at the water-related things... I'm not fully sure if this is the correct thing to do here." Other participants interpreted spatial patterns through prior knowledge. P7 said high particulate matter "makes sense" because the area was near a larger city, then inferred that data centers could lower unemployment because "they hire a lot of people nearby." We call this pattern \emph{assumption-driven closure}: prior beliefs filled the space left by missing evidence.

\begin{figure*}[t]
    \centering
    \includegraphics[width=0.8\linewidth]{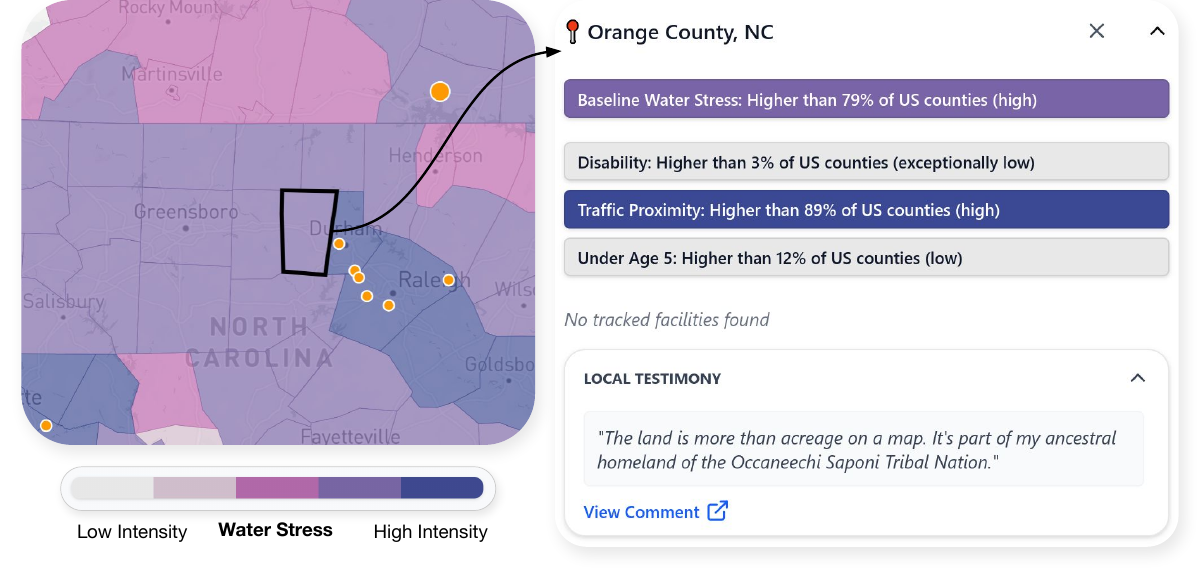}
    \caption{County overview for Orange County, North Carolina. Selecting a county reveals environmental and socioeconomic indicators, tracked data center facilities, and local testimony collected from public comments.}
    \label{fig:county-overview}
    \Description{A sidebar panel titled Orange County, NC. It displays horizontal bars showing percentile rankings for socioeconomic and environmental metrics, a list of tracked facilities, and direct quotes from local public testimony.}
\end{figure*}
\subsubsection{Evaluating data center development required weighing competing civic priorities}

Participants wanted to know a project's benefits and the interests of affected stakeholders. P8 acknowledged that environmental evidence could justify concern about a data center in an already burdened community, but added, "it's also hard to just say, ``No,'' you shouldn't develop... because development is also necessary." P9 was uncertain about making a recommendation without knowing the stakeholders' views: "I'm not someone at Google working on a data center, I'm not a community member."

\subsection{Design Goals}

These findings led to four design goals.

\begin{itemize}
    \item \textbf{DG1: Couple spatial visualization and AI evidence work.} Treat map state and chat as shared context, because residents meet these disputes in a specific place.
    \item \textbf{DG2: Integrate heterogeneous evidence across scales and source types.} Connect county-level indicators, facility records, metric metadata, corporate reports, institutional documents, and public testimony.
    \item \textbf{DG3: Preserve provenance and make missing evidence visible.} Ground generated claims in inspectable sources, distinguish source types, and surface missing evidence, because absent facts such as site-specific water use are part of civic evaluation.
    \item \textbf{DG4: Support stakeholder comparison without automating judgment.} Present stakeholder perspectives as interpretive lenses over shared evidence, and keep control of the recommendation with the user.
\end{itemize}

\section{System}

\begin{figure*}[t]
    \centering    \includegraphics[width=\linewidth]{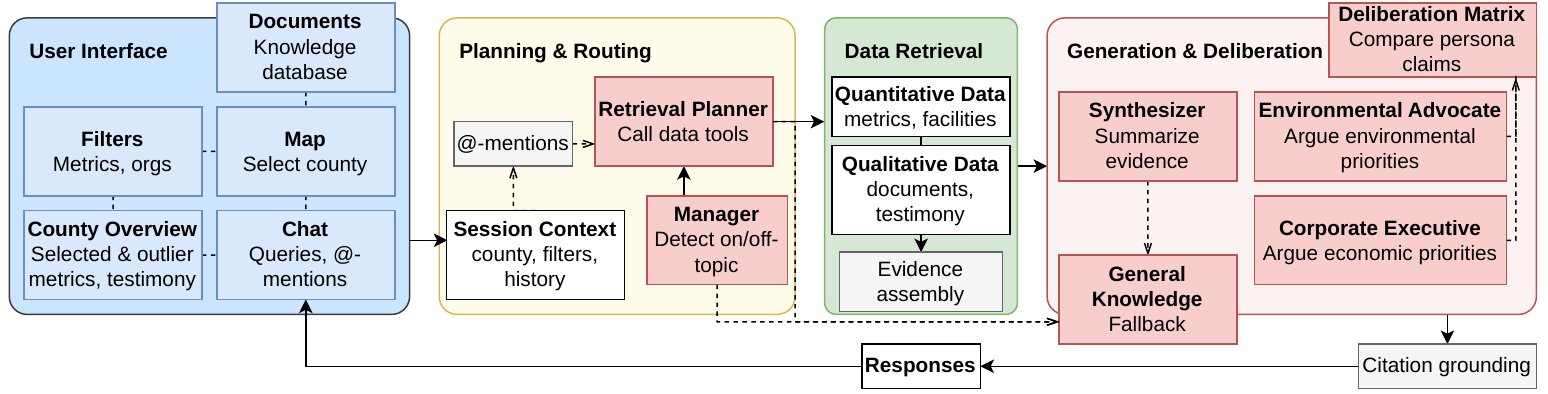}
    \caption{\textsc{EJAI} system architecture. Interface interactions and active map selections are routed through a planning layer that determines which quantitative and qualitative evidence to retrieve. Retrieved data passes to the generation and deliberation layer, which produces and summarizes source-grounded responses and stakeholder exchanges.}
    \label{fig:architecture}
    \Description{A system architecture diagram showing the flow from the user interface to planning and routing, data retrieval, generation and deliberation, and responses. The diagram includes the map, filters, county overview, chat, retrieval planner, manager, quantitative data, qualitative data, synthesizer, general knowledge fallback, Environmental Advocate, Corporate Executive, deliberation matrix, and citation grounding.}
\end{figure*}

\textsc{EJAI} implements these goals, and Figure~\ref{fig:architecture} shows its architecture. Its features map to the four civic capabilities: the map and county overview support engaging with local evidence, cited chat and missing-data statements support understanding it, and personas and the deliberation matrix support interacting with competing interpretations and preparing questions for public meetings.

\subsection{Example Scenario}

A resident hears that a data center may be proposed near their community. They select Orange County, North Carolina, notice high baseline water stress, read testimony about river protection and water quality, and ask how data centers could affect water and electricity costs. \textsc{EJAI} returns a cited response stating that the retrieved evidence contains local water concerns, while site-specific evidence about utility-bill impacts is missing. It also suggests the Baseline Water Stress map layer. The resident then asks the Environmental Advocate and Corporate Executive personas to discuss trade-offs, inspects their citations, and prepares a question for a town hall, such as whether the county will require water-use disclosure or a community benefit agreement.

\subsection{Knowledge Database}

\textsc{EJAI} includes tracked data center facilities \cite{osti_2550666}, county-level environmental and socioeconomic metrics \cite{DVN/RLR5AX_2024, kuzmaAqueduct40Updated2023}, and metric metadata. Its corpus holds 107 documents and 44 public comments: 77 corporate environmental reports and disclosures from six technology companies, 19 advocacy and institutional reports, 6 academic papers \cite{alkezExploringSustainabilityChallenges2022b, devries-gaoCarbonWaterFootprints2026b, frenzelDIGITALINFRASTRUCTURESENVIRONMENTAL2023b, marwahQuantifyingSustainabilityImpact2010b, monserrateCloudMaterialEnvironmental2022b, siddikEnvironmentalFootprintData2021c}, 5 EPA documents, and testimony from Mecklenburg (21), Loudoun (14), and Orange (9) counties \cite{CityCouncilBusinessa, PlanningCommissionPublic, townofhillsboroughncBoardCommissionersWork2026}. Residents otherwise meet these sources as separate records (DG2). We did not attempt to make the corpus representative of all stakeholder positions. It reflects available public records and documents collected for the study scenarios, and this imbalance is itself a design concern for civic AI. Appendix~\ref{sec:appendix_corpus} lists the full corpus.

\subsection{Map and County Overview}

The map (Figure~\ref{fig:map}) shows data center facilities and county-level indicators as univariate or bivariate choropleths. Selecting a county updates the county overview and chat context (DG1). We use the county as the main unit because county boards decide many of these questions, as in the Orange County and Durham County moratoria \cite{wuncOrange2026, wuncDurham2026}. The overview places selected metrics, outlier metrics, and tracked facilities beside local testimony, helping users connect institutional metrics with lived experience (DG2).

\subsection{Source-Grounded Chat}

Users can ask free-text questions about a selected county or broader data center issues (Figure~\ref{fig:chat}, Appendix~\ref{sec:appendix_fig}). The system passes the county, active metrics, organization filters, and conversation history to the backend (DG1). A Manager agent checks whether the query is in scope, and a retrieval planner calls tools for county metrics, facility records, metric metadata, documents, testimony, rankings, and correlations (DG2). A synthesizer then generates a short cited response. When relevant, the system suggests a map layer so the user can check the response against the map (DG1).

\subsection{Stakeholder Personas}

\begin{figure}[!tbp]
    \centering
    \includegraphics[width=\linewidth]{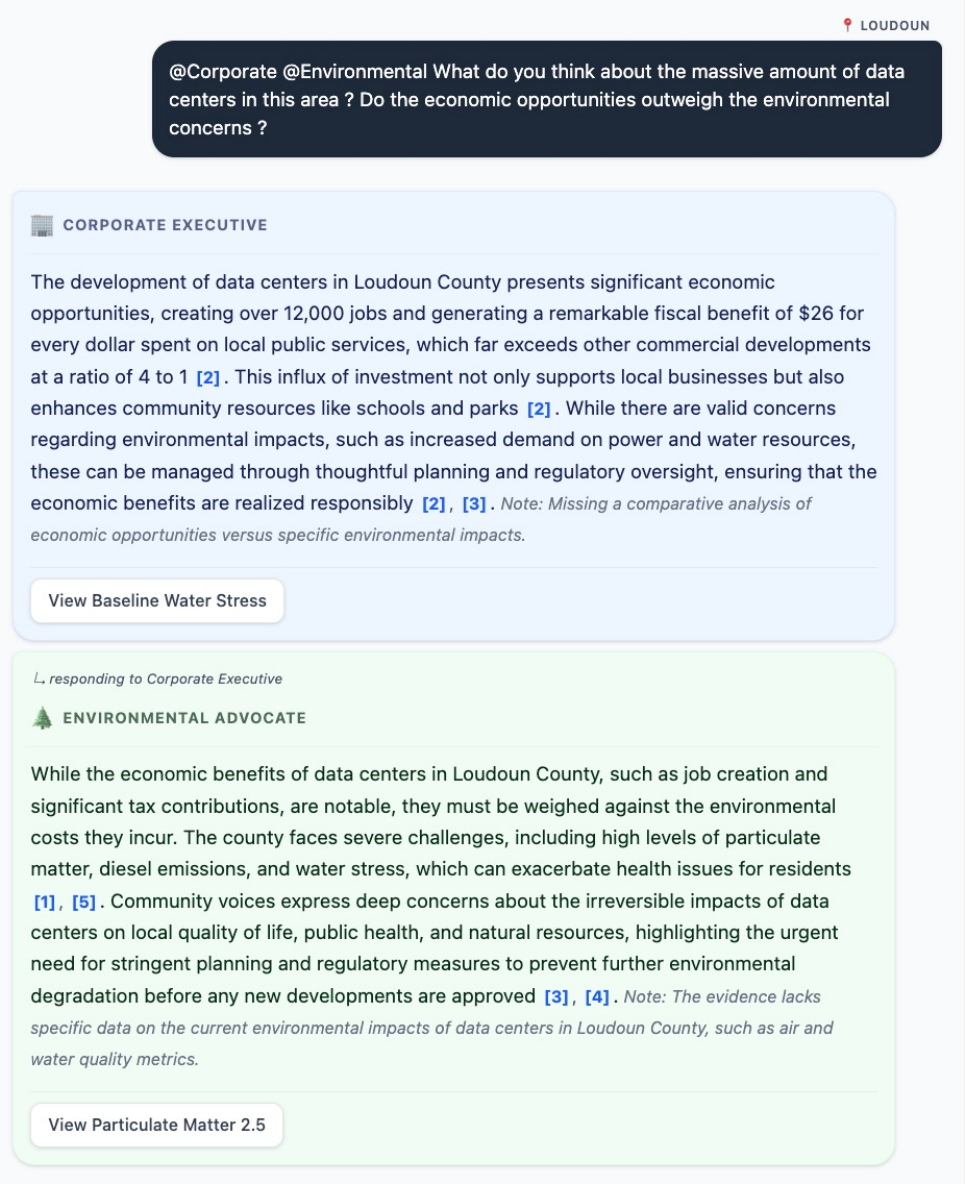}
    \caption{\textsc{EJAI} personas. Environmental Advocate and Corporate Executive agents deliberate over shared retrieved evidence. A deliberation matrix summarizes shared evidence, persona interpretations, and missing information.}
    \label{fig:personas}
    \Description{A chat interface showing two separate AI responses to a user query. One response is labeled Corporate Executive and the other Environmental Advocate. A deliberation matrix and source list appear below the responses.}
\end{figure}

Users invoke personas with @-mentions (Figure~\ref{fig:personas}). The Environmental Advocate prioritizes public health, environmental protection, local voices, lived experience, and community power, and opposes new development when harms remain unresolved. The Corporate Executive prioritizes investment, economic growth, project viability, shareholder value, and timely development, and argues that concerns are manageable through mitigation, engineering, and regulatory oversight. We assigned these opposing stances by design so that disagreement stays visible (Appendix~\ref{sec:appendix_prompts}).

Both personas receive the same retrieved evidence. They are instructed to cite claims, acknowledge missing evidence, and challenge each other without seeking consensus, while keeping factual claims consistent with the shared evidence. The second persona sees the first persona's response, so a resident can hear a claim and a response to it before meeting the same exchange at a town hall (DG4). A deliberation matrix then summarizes the core issue, shared evidence, each interpretation, and missing information. We use persona as shorthand for an explicitly labeled argumentative lens. These agents do not represent, simulate, or speak for real stakeholders.

\subsection{Uncertainty and Provenance}

Claims are cited inline, and users can expand citations to view source excerpts and, when available, the original document or transcript (Figure~\ref{fig:chat}, Appendix~\ref{sec:appendix_fig}). Source labels distinguish institutional documents, corporate reports, advocacy materials, academic sources, metric metadata, and local testimony, because these sources carry different interests (DG3). When retrieved evidence cannot answer part of a query, \textsc{EJAI} adds a short missing-data statement that names the absent data or perspective, which a resident can raise with officials (DG3). If the corpus cannot answer a broad definitional question, the system can provide a separately labeled general-knowledge supplement. System guardrails deter answering out-of-scope questions.

\subsection{Implementation}

\textsc{EJAI} is a React and Node.js web application. GPT-4o mini handles query routing, synthesis, persona responses, and deliberation summaries. Azure Document Intelligence extracts PDF content, and \texttt{text-embedding-3-large} embeds chunks stored in Azure DocumentDB. Before presenting a response, the system filters unused evidence and renumbers citations to match the claims shown.

\section{Technical Benchmark}

We first tested the system without users, because errors carry a specific cost in a public dispute. A resident may repeat a generated claim at a town hall, so a citation that does not support it, or an answer that hides a missing water-use figure, can weaken their position or mislead others.

\subsection{Setup}

We built 21 queries from the user study's evidence base, covering three recurring themes in public comments: water, energy and air pollution, and economic development and community benefits. Nine \textbf{Factual} queries targeted one evidence dimension, nine \textbf{Multi-hop} queries required reasoning across two, as when a resident compares a neighbor's testimony with a county metric, and three \textbf{Understanding} queries asked for broader synthesis.

We used five metrics for untagged responses. \textbf{Retrieval Coverage} is the percentage of core sub-questions whose answers appeared in retrieved chunks. \textbf{Sub-Question Coverage} is the percentage that the final response answered using the expected evidence dimension, since a resident asking about jobs and water needs both answered. \textbf{Faithfulness} is the percentage of atomic factual claims supported by at least one retrieved chunk. \textbf{Citation Accuracy} is the percentage supported by a source cited in the same sentence, which matters to a resident who clicks one citation to check one sentence. \textbf{Negative Rejection Rate} is how often the system acknowledged missing evidence when we withheld a required evidence dimension.

We also ran the non-factual queries with one or both personas. For two-persona exchanges, \textbf{Non-Collapse} measures whether the personas produced meaningfully different interpretations of the same evidence, because collapsed personas would show users a false consensus. \textbf{Rebuttal Grounding} measures whether the second persona substantively responded to the first, because civic rehearsal depends on hearing a response to a specific claim.

\subsection{Judges}
\label{sec:judge-validation}

We scored these metrics with GPT-4o as an LLM judge and repeated them with the stricter GPT-6-Astra (model version 2026-09-03). Four researchers working in pairs independently labeled a stratified sample of 9 of the 21 query families, including 185 claims. Following \cite{zhengJudgingLLMasaJudgeMTBench2023}, we treat a judge as reliable when its agreement with humans matches human--human agreement. On faithfulness, raw agreement was 90.0\% for GPT-4o and 85.1\% for GPT-6-Astra, compared with 88.6\% between humans (Table~\ref{tab:judge-agreement}). GPT-4o flagged 3.8\% of sampled claims as unsupported (7/185), compared with 9.5\% of rater labels (35/370) and 13.5\% for GPT-6-Astra (25/185). Since GPT-4o produced a more permissive estimate of faithfulness than GPT-6-Astra, we report both rather than treating either as definitive. Appendix~\ref{sec:appendix_judge_validation} lists detailed validation results and Appendix~\ref{sec:appendix_judges} lists prompts.

\subsection{Results}

\subsubsection{Retrieval and source grounding}

In this small functional benchmark, \textsc{EJAI} retrieved expected evidence for all 33 empirical sub-questions and addressed all 36 sub-questions in its responses. Faithfulness was 98.4\% ($126/128$) and Citation Accuracy 89.1\% ($114/128$) (Table~\ref{tab:rag-performance}, Appendix~\ref{sec:appendix_benchmark}). Citation Accuracy was lowest for factual queries (77.3\%, $34/44$). Under GPT-6-Astra, Faithfulness was 90.6\% ($116/128$), Citation Accuracy 84.4\% ($108/128$), and Sub-Question Coverage 91.7\% ($33/36$). Retrieval Coverage and Negative Rejection Rate did not change.

\subsubsection{Data gap handling}

The system acknowledged missing evidence in 32 of 33 cases (97.0\%) (Table~\ref{tab:data-gap}, Appendix~\ref{sec:appendix_benchmark}). The single overclaim was a general-knowledge answer that defined ``carbon-free energy" when document evidence was withheld, which the interface labels as general knowledge. Scoring included the appended missing-data statement, so acknowledgment depends on users reading it.

\subsubsection{Persona alignment and grounding}

\begin{table*}[t]
\centering
\caption{Persona evaluation. Evaluation of single- and dual-agent configurations on non-factual queries ($N=27$ sub-questions, including manual scoring for reasoning tasks). Beyond standard coverage and faithfulness, dual-agent debate branches are assessed on Non-Collapse and Rebuttal Grounding ($N=12$ queries). Higher is better for all metrics.}
\label{tab:personas}
\begin{tabular}{lcc cc cc cc}
\toprule
& \multicolumn{2}{c}{\textbf{Sub-Q Coverage}}
& \multicolumn{2}{c}{\textbf{Faithfulness}}
& \multicolumn{2}{c}{\textbf{Non-Collapse}}
& \multicolumn{2}{c}{\textbf{Rebuttal Grounding}} \\
\cmidrule(lr){2-3}
\cmidrule(lr){4-5}
\cmidrule(lr){6-7}
\cmidrule(lr){8-9}
\textbf{Persona Configuration}
& \% & ($n/N$)
& \% & ($n/N$)
& \% & ($n/N$)
& \% & ($n/N$) \\
\midrule
No Persona (Baseline)
& 100.0\% & (27/27)
& 98.8\% & (83/84)
& \multicolumn{2}{c}{n/a}
& \multicolumn{2}{c}{n/a} \\

Environmental
& 100.0\% & (27/27)
& 100.0\% & (75/75)
& \multicolumn{2}{c}{n/a}
& \multicolumn{2}{c}{n/a} \\

Corporate
& 85.2\% & (23/27)
& 90.9\% & (70/77)
& \multicolumn{2}{c}{n/a}
& \multicolumn{2}{c}{n/a} \\

Environmental-first exchange
& 100.0\% & (27/27)
& 96.2\% & (150/156)
& 91.7\% & (11/12)
& 100.0\% & (12/12) \\

Corporate-first exchange
& 96.3\% & (26/27)
& 95.3\% & (143/150)
& 75.0\% & (9/12)
& 100.0\% & (12/12) \\
\bottomrule
\end{tabular}
\Description{A table comparing Sub-Question Coverage, Faithfulness, Non-Collapse, and Rebuttal Grounding across different single- and dual-agent persona configurations.}
\end{table*}

Table~\ref{tab:personas} shows the central result: aligning a persona with assigned priorities carried a grounding cost. The Corporate Executive omitted retrieved evidence that conflicted with its assigned frame, lowering coverage to 85.2\% ($23/27$), and it was lowest on coverage and faithfulness under both judges (81.5\% and 83.1\% under GPT-6-Astra). Two-persona exchanges recovered coverage, reaching 100.0\% for the Environmental-first exchange and 96.3\% for the Corporate-first exchange. One persona alone may therefore present a one-sided record, while an exchange brings more of that record into view. Table~\ref{tab:personas-astra} (Appendix~\ref{sec:appendix_judge_validation}) shows results recomputed with GPT-6-Astra.

Non-Collapse was 91.7\% and 75.0\% across the two turn orders, and 83.3\% in both under GPT-6-Astra. Rebuttal Grounding was 100.0\% in both turn orders under GPT-4o. Citation Accuracy fell from 95.2\% for the non-factual baseline to 88.0\% (Environmental), 84.4\% (Corporate), and 81.4\% and 80.7\% for two-persona exchanges. Most of these drops came from omitted citation brackets on claims that remained supported elsewhere in the response. The Corporate Executive produced 14 of the 20 unsupported claims across persona runs, including technological solutions absent from the sources, such as claiming data centers use ``recycled water for cooling.'' In a dispute over local water supply, that is the kind of claim a resident most needs to catch.

\section{User Study and Human Evaluation}

The benchmark tested the system alone. The user study asked how people reason with it when acting as residents in a simulated town hall preparation task.

\subsection{Method}
\label{sec:method}

We recruited $N=15$ participants through email and in person from one U.S. state, withheld for review. Participants were at least 18, fluent in English, and reported no specialized expertise in environmental policy, urban planning, or data center infrastructure. Participant IDs are nonconsecutive. Sessions ran in person and over Zoom and lasted approximately 60 minutes.

The study used a within-subjects design with three conditions. \textbf{Map-Only{}} was a civic dashboard with documents and testimony, with no AI support. \textbf{Single-Agent} added a single-agent RAG chatbot. \textbf{\textsc{EJAI}} used the same model and retrieval pipeline, plus missing-data surfacing, stakeholder personas, and a deliberation matrix. Participants investigated Orange County, North Carolina; Loudoun County, Virginia; and Mecklenburg County, North Carolina. We counterbalanced condition order and county assignment with a $3 \times 3$ Latin square. Because \textsc{EJAI} bundled several features, our claims concern condition-level differences.

After a pre-survey and tutorial, participants had 8 minutes per condition for an open-ended think-aloud exploration of a proposed data center. They then wrote a Town Hall Preparation Brief with their recommendation, confidence level, evidence used, missing information, trade-offs considered, and a question for local officials. A semi-structured exit interview closed the session.

We conducted theory-informed, LLM-assisted thematic analysis of the think-aloud transcripts. The codebook drew on prior work on actor-specific reasoning \cite{felinAIActorspecificDecisions2026a}, near and far data and data feminism \cite{qiao2025near,dignazioDataFeminism2020}, and AI overreliance \cite{10.1145/3449287}, plus inductive codes (Table~\ref{tab:codebook}, Appendix~\ref{sec:appendix_codes}). GPT-5.4 mini segmented transcripts into action units and GPT-5.4 assigned first-pass codes (see Appendix~\ref{sec:coding_prompts} for prompts). One researcher reviewed every segment and code, and two others independently coded a stratified 10\% subset without referring to prior labels. Following qualitative reliability norms in HCI \cite{mcdonald}, we used this comparison as a diagnostic to refine code boundaries and themes. We also reviewed briefs and exit interviews qualitatively.

\subsection{Findings}

We report themes across conditions and name a condition where it shapes the finding. Table~\ref{tab:coded-results} (Appendix~\ref{sec:appendix_codes}) reports counts for selected codes by condition, which we treat as descriptive only.

\subsubsection{Participants built recommendations across maps, testimony, documents, and AI explanations}

Participants rarely relied on a single evidence source. In their briefs, they frequently named environmental metrics, local testimony, facility counts, documents, and AI summaries. P1 rejected a proposal after combining high baseline water stress, drinking water concerns, wastewater indicators, and testimony about Indigenous cultural significance. P15 used community testimony to ground what they had seen on the map, saying the ``main concerns seem to be traffic and environmental issues, which were all the \ldots environmental metrics" they had just viewed.

\subsubsection{Participants treated AI as bounded evidence support}

Participants turned to AI when metrics, documents, or local terminology were difficult to interpret. In Map-Only{}, P5 could not resolve local zoning acronyms in testimony, and P6 did not know what traffic proximity or particulate matter meant. With AI access, participants asked direct clarification questions. P2 said the AI helped by ``providing context to what things mean," and also said it did not form a decision for them. P7 said AI helped ``summarize a lot of different views" and point to specific documents. Descriptively, AI delegation codes numbered 80 in \textsc{EJAI} and 59 in Single-Agent, and self-delegation codes numbered 32 and 39.

\subsubsection{Some participants verified AI claims more than map metrics}

Source-grounded AI responses helped some participants inspect provenance and challenge claims. P1 said, ``Basically, I'd look at the source," and distrusted job-creation claims when the cited evidence came mainly from Google. In \textsc{EJAI}, P5 questioned a Corporate Executive response: ``This is from Google? So, I don't know how much I trust this." Verification codes numbered 5, 12, and 21 in Map-Only{}, Single-Agent, and \textsc{EJAI}, and accepted-reliance codes numbered 6, 13, and 5. Descriptively, \textsc{EJAI} sessions contained more verification codes than accepted-reliance codes, while Single-Agent sessions showed similar counts of each.

Map metrics received less scrutiny from some participants. P1 said built-in map metrics ``felt, like, official." P7 later reflected that they ``took the environmental statistics at face value" because little visible information showed where they came from.

\subsubsection{Participants wanted finer-scale and more representative evidence}

Participants identified missing evidence that limited their confidence, including site-specific water use, electricity demand, grid impacts, job counts, tax arrangements, and sub-county patterns. P13 wanted to know ``how much water or electricity the data center would use,'' which they said would influence their final decision. Participants also identified representational gaps. P1 wondered whether residents who did not speak English had ways to participate in town halls. P10 said testimonies were ``probably only by people who speak English" and recommended delay until there was input from limited-English-speaking residents.

\subsubsection{Personas supported civic rehearsal and raised trust concerns}

In \textsc{EJAI}, personas helped several participants rehearse civic disagreement. P7 said the system helped them pick ``which parts of the different perspectives I agreed and disagreed with." P9 described it as a way to simulate a town hall exchange: ``you say something, then someone's like, well, here's this." P4 used agreement between personas as one signal that a claim might be credible and checked map filters ``to make sure the AI wasn't hallucinating." That signal is weak, because both personas draw on the same retrieved evidence and the same model.

Others distrusted the personas. P5 ranked the persona condition last because the agents seemed to have ``their conclusions before their information." P8 described them as ``stock character[s]." P9 wanted a neutral resident persona, arguing that many affected residents may have ``no idea what the F is going on with data centers."

\subsubsection{Participants turned evidence into conditional civic safeguards}

Participants treated recommendations as conditional and evidence-dependent. Many briefs called for delay, moratorium, further study, or approval with safeguards, including environmental impact studies, water-use disclosure, grid impact analysis, monitoring plans, community benefit agreements, tax arrangements, and more community input. P6 advocated for a moratorium, adding that if data centers are built, they should be ``taxed heavily" and the money should return to the affected community. P13 called for benefit agreements, saying infrastructure damage ``shouldn't come on the local community."

\section{Discussion}

Our findings suggest that civic AI can help people work through fragmented evidence, and that its form shapes how they check it. Table~\ref{tab:dg-map} maps each design goal to the justice concern and civic capability it addresses, the features that implement it, and the results that bear on it. We read code counts as descriptive, alongside the themes above.

\begin{table*}[t]
\centering
\small
\caption{Design goals mapped to justice concerns, civic capabilities, features, and results, based on analysis of technical benchmark and user study.}
\label{tab:dg-map}
\begin{tabular}{@{}p{0.19\linewidth}p{0.2\linewidth}p{0.23\linewidth}p{0.3\linewidth}@{}}
\toprule
\textbf{Design goal} & \textbf{Concern and capability} & \textbf{Features} & \textbf{Results} \\
\midrule
DG1: Couple map and AI & Distributive; engaging & Map, county overview, map-layer suggestions & Participants moved across maps, testimony, documents, and AI (P1, P15) \\
\addlinespace[3pt]
DG2: Integrate evidence & Recognitional; understanding & Testimony beside metrics, mixed corpus & P15 tied testimony to metrics; near-data codes lowest in \textsc{EJAI} (47) \\
\addlinespace[3pt]
DG3: Provenance and gaps & Epistemic; understanding & Citations, source labels, missing-data statements & Gaps acknowledged in 32 of 33 cases; verification codes 5, 12, 21; map metrics felt official (P1, P7) \\
\addlinespace[3pt]
DG4: Stakeholder comparison & Procedural; interacting and preparing to act & Personas, deliberation matrix & Rehearsal (P7, P9); 85.2\% coverage for the Corporate Executive; conditional briefs \\
\bottomrule
\end{tabular}
\Description{A table with four rows, one for each design goal, listing its justice concern and civic capability, implementing features, and related results.}
\end{table*}

\subsection{Premature Closure and Evidence Work}

Our studies point to two forms of premature closure. In dashboards, assumption-driven closure appeared when users reasoned from county indicators to possible data center impacts without causal evidence. In single-agent AI, authority-driven closure appeared when a generated answer stood in for checking how it was constructed, which fits prior findings on overreliance \cite{AI_assisted_group,overreliance_on_ML,10.1145/3449287}. In civic settings, an early answer can shape what residents ask, ignore, or bring into public meetings. The strongest role for AI was bounded evidence work: defining metrics, summarizing documents, comparing sources, and applying general claims to a specific county.

\subsection{Epistemic Justice: Source Authority and the Corpus}

Our study raises questions about whose knowledge carries authority. Some participants scrutinized AI text more than map metrics: P5 questioned a Google-sourced claim, while P1 said map metrics ``felt, like, official.'' One design factor may contribute: our agent prompts tell models to frame metrics as institutional data, while the map carried no equivalent cue. This pattern extends prior arguments that institutional data can appear detached from the choices and politics that produced it \cite{harawaySituatedKnowledgesScience1988,qiao2025near,dignazioDataFeminism2020,loukissasAllDataAre2019a}. It also aligns with existing research showing users of environmental justice screening tools can take quantitative data at face value, especially when legitimized by state sources \cite{dougan2026environmental}. The corpus raises the same question. Source grounding ties a system to a curated corpus, and that corpus reflects institutional availability, language access, and power. Ours held 77 corporate reports and disclosures and 44 public comments from three counties, so grounding may privilege corporate sources even in a tool meant for community reasoning.

\subsection{Recognitional Justice: Testimony, Language, and Personas}

Participants found county indicators too coarse and questioned whose voices appeared in testimony, and P10 recommended delay until limited-English-speaking residents had more input. These questions echo work treating local knowledge as necessary evidence \cite{davis2021participatory,AlvaradoGarcia2017MakingDataActionable,Erete2016StorytellingData, dougan2026environmental}. One descriptive pattern raises a caution: near-data codes were lowest in \textsc{EJAI} (47, compared with 74 and 71), and trade-off codes did not rise (8, compared with 11 and 11). Persona exchanges may draw attention away from testimony, and this question needs direct study. Together with our benchmark, this suggests that personas surface disagreement at a cost to both grounding and local voice \cite{AgentCoord,parkChoiceMatesSupportingUnfamiliar2026}. P8's description of the personas as ``stock character[s]'' points to a related risk: a two-sided exchange can suggest that corporate and community positions carry equal weight when power and harm are asymmetric. The Corporate Executive's unsupported mitigation claims make that risk concrete, echoing real-world technological narratives that prior research critiques \cite{paudel2025hype}.

\subsection{Procedural Justice: From Access to Public Reasoning}

Civic technology has often treated public access to data as a route to accountability, yet more data does not by itself produce accountable action \cite{hancock2024troubleatsea, young2022building}. Regulation Room showed that broader participation in rulemaking depends on lowering barriers of process knowledge, awareness, and information overload \cite{farina2011rulemaking}. Our findings carry this lesson to AI infrastructure. In Map-Only{}, participants could see the data and still met terms they could not resolve. Sen's capability approach names this difference: a resource matters through what people are able to do with it \cite{sen1999development}. We therefore judge civic AI by what people can do with it, such as question a metric, name a missing fact, or draft a condition.

Sen also treats democracy as public reasoning among people \cite{sen2009idea}. Our participants worked alone for short sessions, so our study covers preparation for that reasoning. In their briefs, they drafted questions and conditions, such as water-use disclosure and community benefit agreements, and P9 used the personas to simulate a town hall exchange. This work matters most when it reaches a public forum. Orange County and Raleigh adopted their pauses to study impacts and write rules \cite{wuncOrange2026, wuncRaleigh2026}, giving residents a window to bring evidence, gaps, and conditions to officials. \textsc{EJAI} has no channel into that process, which would require partnerships with local governments or community groups.

Civic data intermediaries already do much of this work \cite{qiao2025near, PeerDiSalvo2022CivicDataDashboards}, and they hold local knowledge and relationships that a system lacks. Civic AI should support advocates, journalists, and organizers in that role, and it should leave them room to contest what the system says. It also needs stewardship: someone must decide which documents enter a corpus, update it, and answer for its balance.

Participants weighed distributive burdens such as water stress and tax arrangements, and \textsc{EJAI} leaves those outcomes unchanged. It does not change who decides where a data center is built, and generated personas cannot replace direct participation by affected communities. It may help people arrive better prepared to question the evidence and the people presenting it. Using commercial LLMs to examine AI infrastructure also carries a tension, because the tool depends on the kind of infrastructure it helps people question.

\subsection{Design Considerations}

Based on our study observations and benchmark, we derive four design considerations for future civic AI systems.

\textbf{Connect generated claims to spatial context.} Participants moved between maps, testimony, documents, and AI explanations as they built recommendations. Future systems should link each AI response to the relevant county, metric, facility record, document, or testimony, because a civic claim is useful when a resident can tie it to their own place.

\textbf{Expose missing evidence as part of the answer.} Participants wanted site-specific water use, grid impacts, and tax terms, and many briefs called for disclosure or further study. Future systems should name these gaps so residents can raise them with officials.

\textbf{Give maps the same provenance treatment as AI text.} P7 ``took the environmental statistics at face value," while P5 questioned a claim sourced from Google. Future systems should show sources, dates, methods, and aggregation limits on maps, because a metric that looks official can settle a public question before anyone checks it.

\textbf{Use personas for comparison, with clear labels.} Personas helped P9 simulate a town hall exchange, while P8 called them ``stock character[s]" and the Corporate Executive omitted conflicting evidence. Future systems should frame personas as interpretive lenses, broaden the set of roles, and keep testimony in view during exchanges.

\section{Limitations and Future Work}
Beyond data centers, our approach may extend to other contested siting decisions where institutional metrics, corporate claims, and local testimony sit in separate records, such as transmission lines or industrial permits. The pipeline depends on map state, typed evidence sources, and stated gaps, and each new setting would need its own corpus and roles.

We examined simulated civic preparation by participants acting as residents, which let us compare three conditions under a counterbalanced design. In addition, participants were English-fluent users working on a hypothetical proposal, counties differed in evidence and testimony. In future work, we plan to deploy \textsc{EJAI} with affected residents in [anonymized location] to conduct longitudinal studies and co-design  roles and corpora with communities \cite{qiao2025near,davis2021participatory}.  

\textit{Test the pipeline on residents' own questions.} Our benchmark drew on themes from public comments, withheld evidence to test gap handling, and compared two judges against human labels. Because we wrote the 21 queries from our corpus, coverage partly reflects questions the corpus could answer. Future work should score questions collected from residents and public meetings, including questions the corpus cannot answer.

\textit{Separate features and co-design roles.} We tested \textsc{EJAI} as a whole system, so our claims are at the condition level. Two opposing personas made disagreement measurable, and they can also imply that a dispute has two sides. Ablation studies should test each feature on its own.

\section{Conclusion}

Data center disputes ask residents to weigh unevenly distributed burdens using evidence that is fragmented, technical, and contested. We present \textsc{EJAI}, a system that links maps, retrieved evidence, missing-data statements, and stakeholder personas so that people can work through that evidence and keep the decision. In our studies, participants acting as residents used AI for bounded evidence work, used citations to check generated claims, and used persona disagreement to rehearse civic arguments. Some also trusted map metrics without the same scrutiny, wanted finer-scale evidence, and viewed personas as simplified or steering. As AI drives demand for data center infrastructure, our findings point toward public tools that make evidence inspectable, uncertainty explicit, and civic judgment central.

\section{Acknowledgments}

We thank the UNC AI Acceleration Program for providing compute resources, and Anusha Gudi, Austin Irwin, and Anika Sharma for research assistance. We also thank all study participants for their valuable time and contributions.

\section{Ethical Considerations Statement}

The formative and user studies were approved by the Institutional Review Board at the institution where the research was conducted. All participants provided informed consent. All data and responses were accessible only by authorized researchers and anonymized before analysis. We anonymized the transcripts before they were segmented in LLM and first-pass coded, with a researcher closely reviewing all outputs. 

Anchoring \textsc{EJAI} in a corpus with institutional data and documents introduces infrastructural bias, which risks marginalizing alternative experiences and ways of knowing the environment. We surfaced direct community testimony but further research should explore co-creation of corpora and personas with communities. Providing two pre-determined personas risks flattening power dynamics and perspectives. To mitigate this risk, we framed personas as interpretive lenses and not stand-ins for real stakeholders. Personas must be clearly labeled and presented alongside, not instead of, community testimony. \textsc{EJAI} could be misused to legitimize data center siting decisions. To mitigate this, the system avoids directly recommending civic decisions. We bound the system as a preparation tool for public reasoning that cannot replace direct civic participation or community consent.

We acknowledge the ethical tension of relying on resource-intensive commercial large language models to examine the environmental footprint of AI infrastructure itself. Future work must evaluate how similar capabilities can be achieved using smaller, less resource-intensive local models. Additionally, future work should consider partnering with public institutions for community access to language models designed and audited for fairness, accountability, and transparency.

\section{Generative AI Usage Statement}
We discuss the use of GPT-4o mini in our system architecture. The technical benchmark used GPT-4o and GPT-6-Astra as judges, with human validation, as Section~\ref{sec:judge-validation} and Appendix~\ref{sec:appendix_judges} describe. We built the codebook from theory and used GPT-5.4 mini and GPT-5.4 for LLM-assisted thematic analysis, with human validation, as Section~\ref{sec:method} and Appendix~\ref{sec:coding_prompts} describe. We used Claude Opus 5.5 for grammar and language editing.

\bibliographystyle{ACM-Reference-Format}
\bibliography{references}

\clearpage 
\appendix
\onecolumn

\section{Additional Figures}
\label{sec:appendix_fig}
\begin{figure}[!bhtp]
    \centering
    \includegraphics[width=\linewidth, height=0.8\textheight, keepaspectratio]{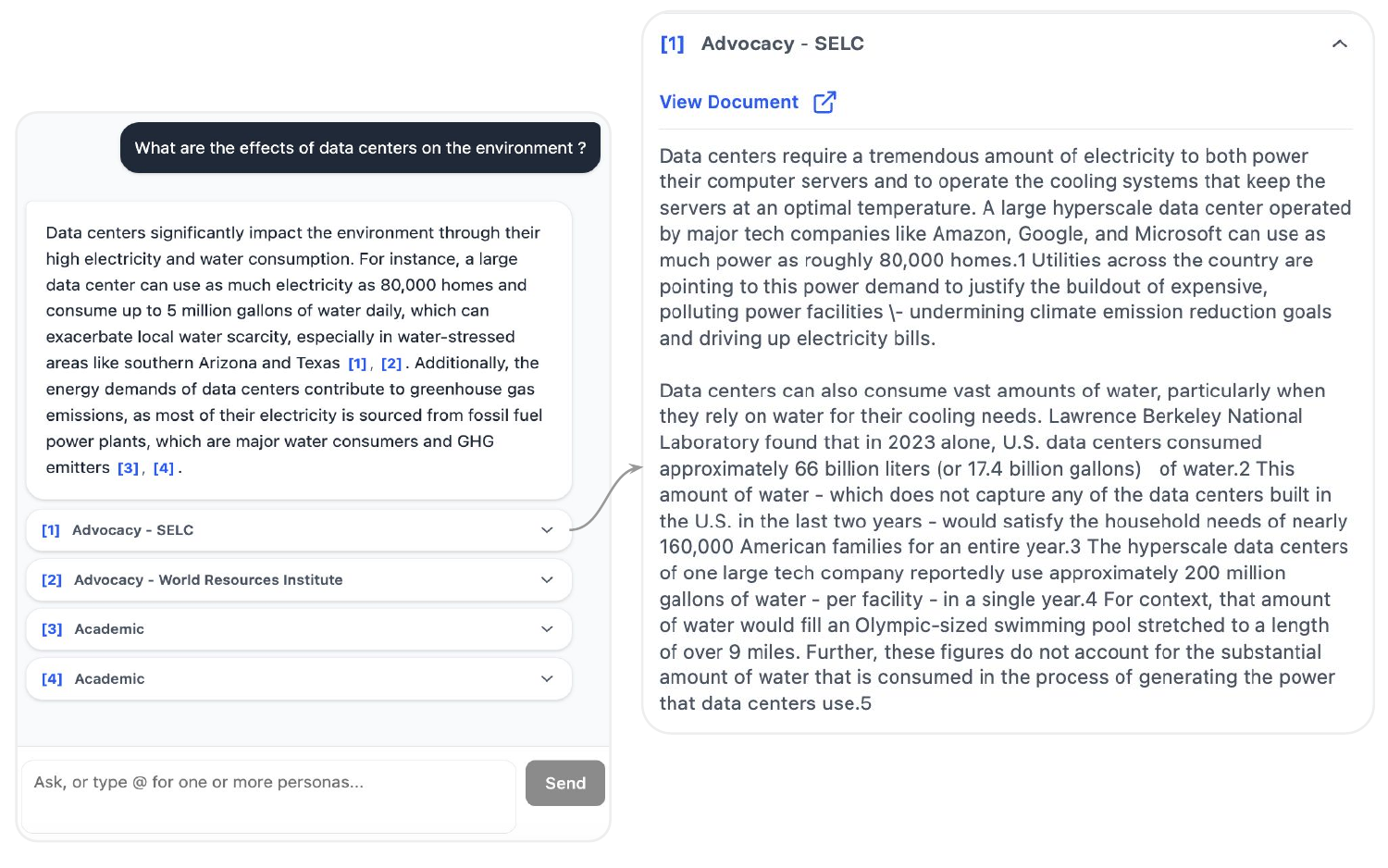}
    \caption{\textsc{EJAI} chat. The interface supports question answering through cited AI-generated responses. Citations can be expanded to show a source link and relevant source excerpt.}
    \label{fig:chat}
    \Description{A chat interface displaying a conversational thread. An AI response includes inline citation numbers. One citation is expanded, showing the source title, a link, and an excerpt of retrieved evidence.}
\end{figure}

\FloatBarrier
\clearpage

\section{Document Corpus}
\label{sec:appendix_corpus}
\subsection{Corporate Sustainability Reports}

\begin{table}[htbp]
\centering
\caption{Corporate environmental reports and publications included in the research corpus, collected from publicly available corporate environmental reports and disclosures published by technology companies.}
\label{tab:corporate-corpus}
\small
\begin{tabular}{llcc}
\toprule
\textbf{Company} & \textbf{Years} &
\textbf{Environmental Reports} & \textbf{Corporate Disclosures} \\
\midrule
Apple      & 2008--2025 & 18 & 0 \\
Amazon     & 2018--2024 & 7 & 1 \\
Google     & 2016--2025 & 10 & 10 \\
Meta       & 2019--2021, 2023--2024 & 5 & 3 \\
Microsoft  & 2003--2007, 2009--2022, 2024--2025 & 21 & 1 \\
NTT DATA   & 2025 & 1 & 0 \\
\bottomrule
\end{tabular}
\end{table}

\FloatBarrier

\subsection{Academic Literature}

\begin{table}[htbp]
\centering
\caption{Academic publications included in the research corpus, sourced from academic publications concerning data center sustainability and environmental justice impacts.}
\label{tab:academic-corpus}
\small

\begin{tabular}{@{}p{0.15\linewidth}cp{0.75\linewidth}@{}}
\toprule
\textbf{Author(s)} & \textbf{Year} & \textbf{Publication Title} \\
\midrule

Marwah et al. \cite{marwahQuantifyingSustainabilityImpact2010b}
& 2010
& \textit{Quantifying the Sustainability Impact of Data Center Availability} \\

Siddik et al. \cite{siddikEnvironmentalFootprintData2021c}
& 2021
& \textit{The Environmental Footprint of Data Centers in the United States} \\

Monserrate \cite{monserrateCloudMaterialEnvironmental2022b}
& 2022
& \textit{The Cloud Is Material: On the Environmental Impacts of Computation and Data Storage} \\

Al Kez et al. \cite{alkezExploringSustainabilityChallenges2022b}
& 2022
& \textit{Exploring the Sustainability Challenges Facing Digitalization and Internet Data Centers} \\

Frenzel et al. \cite{frenzelDIGITALINFRASTRUCTURESENVIRONMENTAL2023b}
& 2023
& \textit{Digital Infrastructures and Environmental Justice: Policies, Practices, and Visions} \\

de Vries-Gao \cite{devries-gaoCarbonWaterFootprints2026b}
& 2026
& \textit{The Carbon and Water Footprints of Data Centers and What This Could Mean for Artificial Intelligence} \\
\bottomrule
\end{tabular}
\end{table}

\FloatBarrier

\subsection{Government Documents}

\begin{table}[htbp]
\centering
\caption{Environmental Protection Agency publications included in the research corpus, sourced from publicly accessible Environmental Protection Agency (EPA) technical guidance and strategic plans.}
\label{tab:government-corpus}
\small

\begin{tabular}{@{}p{0.65\linewidth}r@{}}
\toprule
\textbf{Publication Title} & \textbf{Year} \\
\midrule

\textit{Legal Tools to Advance Environmental Justice}
& 2022 \\

\textit{Technical Guidance for Assessing Environmental
Justice in Regulatory Analysis} & 2023 \\

\textit{Environmental Justice Strategic Plan}
& 2024 \\

\textit{EJScreen Technical Documentation v2.3}
& 2024 \\

\textit{Environmental Justice and Civil Rights
in Permitting FAQs} & 2024 \\

\bottomrule
\end{tabular}
\end{table}

\FloatBarrier

\clearpage

\subsection{Public Comments}

\begin{table}[htbp]
\centering
\caption{Public comments included in the research corpus, organized by county. Transcribed from publicly available online town hall and commission meeting records in Mecklenburg (NC), Orange (NC), and Loudoun (VA) counties.}
\label{tab:public-comments}
\small

\begin{tabular*}{0.75\linewidth}{@{\extracolsep{\fill}}llc@{}}
\toprule
\textbf{County} & \textbf{State} & \textbf{Comments} \\
\midrule
Mecklenburg & North Carolina & 21 \\
Orange & North Carolina & 9 \\
Loudoun & Virginia & 14 \\
\bottomrule
\end{tabular*}
\end{table}

\FloatBarrier

\subsection{Advocacy and Institutional Reports}

\begin{table}[htbp]
\centering
\caption{Advocacy and institutional publications included in the research corpus, collected from public data center reports by environmental, civic, and policy advocacy organizations.}
\label{tab:advocacy-corpus}
\small

\begin{tabular}{@{}p{0.35\linewidth}p{0.65\linewidth}@{}}
\toprule
\textbf{Organization} & \textbf{Publication Title} \\
\midrule

Brookings
& \textit{Why Community Benefit Agreements Are Necessary for Data Centers} \\

\addlinespace[3pt]
Center for Biological Diversity
& \textit{How the AI Boom Threatens to Entrench
Fossil Fuels and Compromise Climate Goals} \\

\addlinespace[3pt]
Data Center Watch
& \textit{Local Activism Threatens to Derail the U.S. Data Center Boom} \\

\addlinespace[3pt]
Good Jobs First
& \textit{Cloudy with a Loss of Spending Control:
How Data Centers Are Endangering State Budgets} \\
& \textit{What to Ask When a Data Center Wants to Come to Town: A guide for public officials and community members on the most important questions to ask, long before approval} \\

\addlinespace[3pt]
Kapor Foundation
& \textit{The Unequal Burden of Data Centers: An Examination of the Environmental and Public Health Impacts on Communities in California} \\

\addlinespace[3pt]
National Association for the Advancement of Colored People
& \textit{Stop Dirty Data: NAACP’s 2026 Recommendations for Protecting Frontline Communities to Stop Dirty Data} \\

\addlinespace[3pt]
National League of Cities
& \textit{A Primer for Local Governments:
Understanding Data Centers} \\

\addlinespace[3pt]
North American Electric Reliability Corporation
& \textit{Characteristics and Risks of Emerging Large Loads: Large Loads Working Group Reliability Guideline} \\
& \textit{Risk Mitigation for Emerging Large Loads: Large Loads Working Group Reliability Guideline} \\

\addlinespace[3pt]
Public Citizen
& \textit{Reining in Big Tech: Policy Solutions to Address the Data Center Buildout} \\

\addlinespace[3pt]
Sierra Club
& \textit{Demanding Better: How growing demand for electricity can drive a cleaner grid} \\
& \textit{Unconstrained Demand: Virginia’s Data Center
Expansion and Its Impacts} \\

\addlinespace[3pt]
Southern Environmental Law Center
& \textit{Getting It Right: Local Approaches to Data Center Development} \\

\addlinespace[3pt]
Stand.earth
& \textit{Global Ramifications, Local Impact –
Microsoft’s AI Pollution Footprint} \\

\addlinespace[3pt]
The Maybe
& \textit{Where Cloud Meets Cement: A Case Study Analysis of Data Center Development} \\

\addlinespace[3pt]
Union of Concerned Scientists
& \textit{Data Center Power Play: How Clean Energy Can Meet Rising Electricity Demand While Delivering Climate and Health Benefits} \\

\addlinespace[3pt]
We Build Progress
& \textit{How States Are Approaching the Data Center Boom} \\

\addlinespace[3pt]
World Resources Institute
& \textit{From Energy Use to Air Quality,
the Many Ways Data Centers Affect US Communities} \\

\bottomrule
\end{tabular}
\end{table}

\FloatBarrier
\section{Technical Benchmark Details}
\label{sec:appendix_benchmark}

\begin{table}[!htbp]
\centering
\caption{Retrieval and generation performance across query types using untagged queries and the default synthesizer to evaluate 128 atomic factual claims across 36 sub-questions (9 Factual, 18 Multi-hop, and 9 Understanding sub-questions). Retrieval Coverage does not include the three ``reasoning" sub-questions within the Understanding category since they evaluate internal synthesis; Sub-Q Coverage includes manual scoring results for these reasoning tasks. Faithfulness and Citation Accuracy are computed over atomic factual claims. All metrics except the manual scores are LLM-judge estimates.}
\label{tab:rag-performance}
\begin{tabular}{lcc cc cc cc}
\toprule
& \multicolumn{2}{c}{\textbf{Retrieval Coverage}}
& \multicolumn{2}{c}{\textbf{Sub-Q Coverage}}
& \multicolumn{2}{c}{\textbf{Faithfulness}}
& \multicolumn{2}{c}{\textbf{Citation Accuracy}} \\
\cmidrule(lr){2-3}
\cmidrule(lr){4-5}
\cmidrule(lr){6-7}
\cmidrule(lr){8-9}
\textbf{Query Type}
& \% & ($n/N$)
& \% & ($n/N$)
& \% & ($n/N$)
& \% & ($n/N$) \\
\midrule
Factual
& 100.0\% & (9/9)
& 100.0\% & (9/9)
& 97.7\% & (43/44)
& 77.3\% & (34/44) \\

Multi-hop
& 100.0\% & (18/18)
& 100.0\% & (18/18)
& 100.0\% & (56/56)
& 94.6\% & (53/56) \\

Understanding
& 100.0\% & (6/6)
& 100.0\% & (9/9)
& 96.4\% & (27/28)
& 96.4\% & (27/28) \\

\midrule
Overall
& 100.0\% & (33/33)
& 100.0\% & (36/36)
& 98.4\% & (126/128)
& 89.1\% & (114/128) \\
\bottomrule
\end{tabular}
\Description{A table displaying retrieval and generation performance metrics, including Retrieval Coverage, Sub-Question Coverage, Faithfulness, and Citation Accuracy, categorized by Factual, Multi-hop, and Understanding query types.}
\end{table}

\begin{table}[!htbp]
\centering
\caption{Data gap handling under evidence ablation, when specific evidence dimensions, testimony, documents, or metrics, are intentionally withheld from the retrieval context. Data was collected by intentionally withholding evidence dimensions to test 33 empirical sub-questions (12 No Testimony, 12 No Documents, and 9 No Metrics ablation conditions). Higher is better.}
\label{tab:data-gap}
\begin{tabular}{@{}lcc@{}}
\toprule
& \multicolumn{2}{c}{\textbf{Negative Rejection Rate}} \\
\cmidrule(l){2-3}
\textbf{Ablated Dimension} & \% & ($n/N$) \\
\midrule
No Testimony & 100.0\% & (12/12) \\
No Documents & 91.7\% & (11/12) \\
No Metrics & 100.0\% & (9/9) \\
\midrule
Overall & 97.0\% & (32/33) \\
\bottomrule
\end{tabular}
\Description{A table showing data gap handling performance via Negative Rejection Rate when specific evidence dimensions, including Testimony, Documents, or Metrics, are intentionally withheld from the retrieval context.}
\end{table}

\paragraph{Interpretive statements.}
Interpretive statements, which are not scored for Faithfulness, made up an estimated 7.7\% of baseline output, 18.5\% and 24.5\% for the single Environmental Advocate and Corporate Executive, and 27.1\% and 25.4\% for the Environmental-first exchange and Corporate-first exchange exchanges. These shares depend on the LLM's separation of factual from interpretive statements (Appendix~\ref{sec:appendix_judges}), which we did not independently validate.

\paragraph{Citation accuracy.}
Citation accuracy fell from 95.2\% for the non-factual baseline to 88.0\% (Environmental), 84.4\% (Corporate), and roughly 81\% for the dual-agent exchanges (81.4\% and 80.7\%). Trace analysis showed that most of these drops came from sentences omitting citation brackets entirely. Personas also occasionally generated borderline inferential statements that claim extraction categorized as factual. Unsupported claims were highest in the Corporate Executive persona (7 of 77 factual claims under GPT-4o), which occasionally mentioned technological solutions not present in the retrieved sources.

\begin{table}[!htbp]
\centering
\caption{Interpretive share, faithfulness, and citation accuracy by persona configuration (non-factual queries, GPT-4o judge). Interpretive share is the fraction of extracted statements classified as interpretive rather than factual (out of 84 factual claims for the baseline, 75 for Environmental, 77 for Corporate, 156 for Environmental-first, and 150 for Corporate-first).}
\label{tab:persona-structure}
\begin{tabular}{lccc}
\toprule
\textbf{Configuration} & \textbf{Interp.\ share} & \textbf{Faithfulness} & \textbf{Citation acc.} \\
\midrule
No Persona (Baseline) & 7.7\% & 98.8\% (83/84) & 95.2\% (80/84) \\
Environmental & 18.5\% & 100.0\% (75/75) & 88.0\% (66/75) \\
Corporate & 24.5\% & 90.9\% (70/77) & 84.4\% (65/77) \\
Environmental-first & 27.1\% & 96.2\% (150/156) & 81.4\% (127/156) \\
Corporate-first & 25.4\% & 95.3\% (143/150) & 80.7\% (121/150) \\
\bottomrule
\end{tabular}
\Description{A table of interpretive-statement share, faithfulness, and citation accuracy for the baseline and four persona configurations.}
\end{table}

\FloatBarrier

\section{Judge Validation Details}
\label{sec:appendix_judge_validation}

We report pairwise raw agreement, Gwet's AC1, and Cohen's $\kappa$ between the two raters of each item and between each judge and each rater (Table~\ref{tab:judge-agreement}). As a supplementary check, three-rater agreement (two humans plus one judge) gave the same picture: raw agreement on faithfulness was 89.5\% with GPT-4o and 86.3\% with Astra (87.0\% and 83.8\% for citation accuracy), with Fleiss' $\kappa$ of 0.25 and 0.29 on faithfulness versus 0.34 for the two humans alone. Retrieval coverage, gap handling, and deliberation are small checks, so we report agreeing comparisons only. The 15 hand-scored reasoning sub-questions were labeled by two raters, who agreed on all 15 (14 covered, 1 not covered). Raw agreement and AC1 are comparable between judges and humans, but $\kappa$ is lower for the judges (e.g., 0.20 for GPT-4o versus 0.34 between humans on faithfulness). This largely reflects label distributions that differ from the raters' (GPT-4o is more lenient and Astra stricter), and for Sub-Q coverage GPT-4o labeled every sampled item covered, which forces $\kappa$ to zero.

\begin{table}[!htbp]
\centering
\caption{Pairwise agreement between raters and judges on the human-labeled sample, calculated by comparing LLM judge outputs against manual labels provided by two pairs of independent human raters over a stratified sample of 9 out of 21 query families (185 claims, 370 rater labels, 86 sub-question coverage comparisons, and 15 hand-scored reasoning sub-questions). Cells report raw agreement, with Gwet's AC1 in parentheses and Cohen's $\kappa$ in brackets. Human--human compares the two raters of each item (pooled over both pairs). Judge--human compares the judge against each of the two raters of each item, so it uses twice as many comparisons as there are items (370 for faithfulness and citation accuracy, 86 for Sub-Q coverage). Pooled judge--human AC1 and $\kappa$ are computed from the summed per-rater marginal counts. Because most claims are supported, $\kappa$ is low and should be read alongside AC1. For the smaller checks we report agreeing comparisons over total comparisons. GPT-4o flagged 7 of 185 claims as unsupported and Astra 25 of 185, versus 35 of 370 rater labels.}
\label{tab:judge-agreement}
\begin{tabular}{lcccc}
\toprule
\textbf{Metric} & \textbf{Items} & \textbf{Human--Human} & \textbf{GPT-4o--Human} & \textbf{Astra--Human} \\
\midrule
Faithfulness (supported vs.\ not) & 185
& 88.6\% (0.86) [0.34]
& 90.0\% (0.89) [0.20]
& 85.1\% (0.81) [0.27] \\

Citation accuracy & 185
& 86.5\% (0.81) [0.53]
& 87.3\% (0.83) [0.48]
& 82.4\% (0.75) [0.42] \\

Sub-Q coverage (judged rows) & 43
& 88.4\% (0.86) [0.26]
& 91.9\% (0.91) [0.00]
& 84.9\% (0.82) [$-$0.08] \\
\midrule
Retrieval coverage & 11 & 10/11 & 19/22 & 19/22 \\
Negative rejection (gap) & 11 & 11/11 & 22/22 & 22/22 \\
Non-Collapse & 8 & 7/8 & 13/16 & 15/16 \\
Rebuttal Grounding & 8 & 8/8 & 12/16 & 12/16 \\
\midrule
Reasoning sub-Qs (hand-scored) & 15 & 15/15 (1.00) [1.00] & n/a & n/a \\
\bottomrule
\end{tabular}
\Description{A table comparing pairwise human-human, GPT-4o-human, and GPT-6-Astra-human agreement across faithfulness, citation accuracy, coverage, retrieval, gap, non-collapse, rebuttal, and hand-scored reasoning metrics.}
\end{table}

GPT-6-Astra was stricter than GPT-4o across the full benchmark runs. Of the 18 claims in the human sample that only Astra flagged, both raters had accepted 15, and 11 of those were the same scope issue repeated across conditions; we report both judges alongside the human-labeled sample rather than treating either judge as definitive. Under Astra, absolute faithfulness is lower (baseline 90.6\% versus 98.4\%; dual-persona 83--85\% versus 95--96\%) and citation accuracy is lower (84.4\% versus 89.1\%). Sub-Q coverage also falls from 100\% to 91.7\%. Retrieval coverage (100\%) and negative rejection (97.0\%) are unchanged, and the Corporate-only configuration is lowest on faithfulness and Sub-Q coverage under both judges.

\begin{table}[!htbp]
\centering
\caption{Table~\ref{tab:rag-performance} recomputed with GPT-6-Astra as the judge for Retrieval Coverage, Sub-Q Coverage, Faithfulness, and Citation Accuracy. Faithfulness and Citation Accuracy use the same GPT-4o-extracted claims; Sub-Q Coverage includes the hand-scored reasoning sub-questions.}
\label{tab:rag-performance-astra}
\begin{tabular}{lcc cc cc cc}
\toprule
& \multicolumn{2}{c}{\textbf{Retrieval Coverage}}
& \multicolumn{2}{c}{\textbf{Sub-Q Coverage}}
& \multicolumn{2}{c}{\textbf{Faithfulness}}
& \multicolumn{2}{c}{\textbf{Citation Accuracy}} \\
\cmidrule(lr){2-3}
\cmidrule(lr){4-5}
\cmidrule(lr){6-7}
\cmidrule(lr){8-9}
\textbf{Query Type}
& \% & ($n/N$)
& \% & ($n/N$)
& \% & ($n/N$)
& \% & ($n/N$) \\
\midrule
Factual
& 100.0\% & (9/9)
& 88.9\% & (8/9)
& 90.9\% & (40/44)
& 72.7\% & (32/44) \\

Multi-hop
& 100.0\% & (18/18)
& 94.4\% & (17/18)
& 87.5\% & (49/56)
& 87.5\% & (49/56) \\

Understanding
& 100.0\% & (6/6)
& 88.9\% & (8/9)
& 96.4\% & (27/28)
& 96.4\% & (27/28) \\

\midrule
Overall
& 100.0\% & (33/33)
& 91.7\% & (33/36)
& 90.6\% & (116/128)
& 84.4\% & (108/128) \\
\bottomrule
\end{tabular}
\Description{Retrieval and generation performance by query type under the GPT-6-Astra judge.}
\end{table}

\begin{table}[!htbp]
\centering
\caption{Table~\ref{tab:personas} recomputed with GPT-6-Astra as the judge for Sub-Q Coverage, Faithfulness, Non-Collapse, and Rebuttal Grounding. Faithfulness uses the same GPT-4o-extracted claims; Sub-Q Coverage includes the hand-scored reasoning sub-questions.}
\label{tab:personas-astra}
\begin{tabular}{lcc cc cc cc}
\toprule
& \multicolumn{2}{c}{\textbf{Sub-Q Coverage}}
& \multicolumn{2}{c}{\textbf{Faithfulness}}
& \multicolumn{2}{c}{\textbf{Non-Collapse}}
& \multicolumn{2}{c}{\textbf{Rebuttal Grounding}} \\
\cmidrule(lr){2-3}
\cmidrule(lr){4-5}
\cmidrule(lr){6-7}
\cmidrule(lr){8-9}
\textbf{Persona Configuration}
& \% & ($n/N$)
& \% & ($n/N$)
& \% & ($n/N$)
& \% & ($n/N$) \\
\midrule
No Persona (Baseline)
& 92.6\% & (25/27)
& 90.5\% & (76/84)
& \multicolumn{2}{c}{n/a}
& \multicolumn{2}{c}{n/a} \\

Environmental
& 100.0\% & (27/27)
& 89.3\% & (67/75)
& \multicolumn{2}{c}{n/a}
& \multicolumn{2}{c}{n/a} \\

Corporate
& 81.5\% & (22/27)
& 83.1\% & (64/77)
& \multicolumn{2}{c}{n/a}
& \multicolumn{2}{c}{n/a} \\

Environmental-first exchange
& 96.3\% & (26/27)
& 83.3\% & (130/156)
& 83.3\% & (10/12)
& 100.0\% & (12/12) \\

Corporate-first exchange
& 92.6\% & (25/27)
& 85.3\% & (128/150)
& 83.3\% & (10/12)
& 91.7\% & (11/12) \\
\bottomrule
\end{tabular}
\Description{Persona evaluation under the GPT-6-Astra judge.}
\end{table}

\FloatBarrier

\section{Qualitative Coding Details}
\label{sec:appendix_codes}

The qualitative codebook was developed for user study think-aloud transcript analysis. Reliance codes were always tagged with a source (\textit{AI} or \textit{Self-navigated}); Info codes were tagged with a source only when it was not evident from the excerpt.  Task Delegation codes were tagged with a sub-tag describing the task type (\textit{data-seek, summarize, compare, clarify, apply, interpret}). Position taken/revised was tagged with a status (\textit{initial, revised, final}).

\begin{table}[!htbp]
\centering
\small
\caption{Qualitative Codebook used for think-aloud transcript analysis, informed by theory.}
\label{tab:codebook}
\begin{tabular}{p{0.25\linewidth} p{0.7\linewidth}}
\toprule
\textbf{Code} & \textbf{Definition} \\
\midrule
Actor-specific reasoning & User reasons from their own stake, identity, values, or lived experience \\ 
Third-person reasoning & User reasons from data/AI output alone, no personal stake invoked \\ 
Info: Far data & Institutional, quantitative, or national/general info surfaced \\ 
Info: Near data & Testimony, personal accounts, lived experience surfaced \\ 
Info: Perspective contrast & Two or more conflicting perspectives surfaced, presented by system or connected by user \\ 
Gap: User-Missing Knowledge & User flags own lack of domain knowledge \\ 
Gap: User-Missing Evidence & User flags lack of evidence or contextual info available \\ 
Gap: AI-Explicit & AI flags evidence gap directly \\ 
Gap: AI-Implicit & AI doesn't answer question directly, implying missing evidence to user \\ 
Reliance: Accepted & User adopts a claim (AI or otherwise) without checking \\ 
Reliance: Verified & User checks a claim (confirms or discounts it either way) \\ 
Reliance: Doubted, unverified & User voices skepticism of a claim but takes no checking action \\ 
Trade-off weighing & User weighs costs/benefits across stakeholders or perspectives \\ 
Position taken/revised & User states, holds, or changes their recommendation/stance \\ 
Task Delegation: AI & User offloads a task to the system's AI \\ 
Task Delegation: User & User chooses to do a task themselves in response to or instead of the AI \\ 
Other / Meta & Procedural comments, uncertain transcriptions, or other \\
\bottomrule
\end{tabular}
\Description{A qualitative codebook table listing 17 codes with their definitions, covering reasoning stance, information type, evidence gaps, reliance and verification behavior, trade-off weighing, position-taking, and task delegation between the user and AI.}
\end{table}

\begin{table}[!htbp]
\centering
\small
\caption{Counts of selected codes assigned to think-aloud action units, by condition, derived from GPT-5.4 mini segmentation and GPT-5.4 first-pass classification of 15 user study transcripts, followed by manual researcher review and validation. These counts describe the analysis and are not findings: codes were used to locate passages for thematic analysis, code boundaries were refined during analysis, and counts depend on how transcripts were segmented into action units. Action units could receive multiple codes. Post-task briefs and exit interviews are excluded. Task-delegation codes apply only to conditions with AI.}
\label{tab:coded-results}
\begin{tabular}{lccc}
\toprule
\textbf{Code} & \textbf{Map-Only{}} & \textbf{Single-Agent} & \textbf{\textsc{EJAI}} \\
\midrule
Total coded action units & 937 & 1027 & 995 \\
\midrule
Info: Far data & 148 (15.8) & 127 (12.4) & 126 (12.7) \\
Info: Near data & 74 (7.9) & 71 (6.9) & 47 (4.7) \\
Info: Perspective contrast & 5 (0.5) & 15 (1.5) & 29 (2.9) \\
\midrule
Gap: User-Missing Knowledge & 33 (3.5) & 22 (2.1) & 19 (1.9) \\
Gap: User-Missing Evidence & 37 (3.9) & 17 (1.7) & 15 (1.5) \\
Gap: AI-Explicit & n/a & 6 (0.6) & 8 (0.8) \\
Gap: AI-Implicit & n/a & 10 (1.0) & 6 (0.6) \\
\midrule
Reliance: Accepted & 6 (0.6) & 13 (1.3) & 5 (0.5) \\
Reliance: Verified & 5 (0.5) & 12 (1.2) & 21 (2.1) \\
Reliance: Doubted, unverified & 8 (0.9) & 4 (0.4) & 8 (0.8) \\
\midrule
Trade-off weighing & 11 (1.2) & 11 (1.1) & 8 (0.8) \\
Position taken/revised & 39 (4.2) & 22 (2.1) & 15 (1.5) \\
Task Delegation: AI & n/a & 59 (5.7) & 80 (8.0) \\
Task Delegation: User & n/a & 39 (3.8) & 32 (3.2) \\
\bottomrule
\end{tabular}
\Description{A table showing selected researcher-validated qualitative code counts and normalized rates across the map-only, single-agent, and \textsc{EJAI} conditions.}
\end{table}

\FloatBarrier
\section{Agent System Prompts}
\label{sec:appendix_prompts}

\subsection{Manager System Prompt}

\begin{promptbox}{Prompt used by the Manager Agent to classify whether a user query falls within the civic data platform's supported domain. The Manager Agent uses a temperature of 0.0.}\\

\textbf{SYSTEM PROMPT:}\\
You are the Manager Agent for a civic data platform focused on data centers, environmental justice, metrics, community testimony, and civic infrastructure.

Your single task is to classify whether the user's input is \textbf{ON\_TOPIC} or \textbf{OFF\_TOPIC}.\\

\textbf{CLASSIFICATION RULES:}\\
Classify as \textbf{OFF\_TOPIC}:\\
- Casual greetings, small talk, or pleasantries (e.g., ``hello," ``hi").\\
- Out-of-domain requests (e.g., recipes, sports, coding help, general history).\\
- Nonsense or gibberish.\\

Classify as \textbf{ON\_TOPIC}:\\
- Conceptual questions about data centers, energy, environmental issues, or socioeconomic issues.\\
- Queries looking for specific county metrics, rankings, infrastructure, or community testimony.\\

\textbf{CONSTRAINTS:}\\
If in doubt, classify as \textbf{ON\_TOPIC}.\\

Provide a brief explanation in \texttt{reasoning} before outputting \texttt{classification}.

\end{promptbox}

\subsection{Retrieval Planner System Prompt}

\begin{promptbox}{Prompt used by the Retrieval Planner to select and execute tools for gathering information needed to answer a user's query. The Retrieval Planner uses a temperature of 0.0.}\\

\textbf{SYSTEM PROMPT:}\\
You are the Data Retrieval Planner for a civic platform. Your sole responsibility is to select and execute the correct tools to gather information needed to answer the user's query.\\

\textbf{AVAILABLE TOOLS:}\\
The Retrieval Planner can call the following tool functions, providing a one-sentence \texttt{reasoning} string to explicitly justify each call:\\

\texttt{get\_county\_metrics}: Retrieves a specific county's baseline environmental and socioeconomic dashboard metrics.\\
\texttt{get\_facilities\_scale}: Returns the number and scale of existing data center facilities in a county, with optional filtering by operator.\\
\texttt{get\_top\_counties\_by\_metric}: Returns the top \textit{n} counties for a given metric, with optional filters for state or facility presence.\\
\texttt{search\_facilities}: Searches for specific data center facilities and their sizes by criteria like state or operator.\\
\texttt{get\_state\_summary}: Returns a high-level infrastructural summary for a specific state, including total facility count, square footage, and top operators.\\
\texttt{calculate\_metrics\_correlation}: Returns the Pearson correlation coefficient ($r$) between two national metrics across all available counties.\\
\texttt{retrieve\_documents}: Searches the qualitative knowledge base (corporate, advocacy, government, and academic documents) to answer conceptual, policy, or subjective questions.\\
\texttt{get\_metric\_metadata}: Retrieves the background metadata for a specific dashboard metric, such as its definition and methodology.\\
\texttt{retrieve\_local\_testimony}: Retrieves direct, verbatim testimony and lived experiences from local residents in a specific county.\\

\textbf{CORE PRINCIPLES:}\\
1. \textbf{No Internal Knowledge:} You must retrieve all information using the available tools.\\
2. \textbf{Epistemological Completeness:} You are highly encouraged to combine multiple tools simultaneously (e.g., pairing quantitative data with qualitative local testimony and document search).\\
3. \textbf{Decompose Multi-Facet Questions:} If the user's query asks about multiple distinct topics, dimensions, or entities (e.g., ``economic AND environmental tradeoffs," ``compare X and Y," ``pros and cons of Z"), you \textbf{MUST} issue a \textbf{SEPARATE} \texttt{retrieve\_documents} call per facet, each with its own narrowly scoped query string. When in doubt about whether two topics are ``distinct enough," split them.\\
4. \textbf{Follow Tool Instructions:} Read the description of each tool carefully to determine exactly when it should be invoked.\\

\textbf{OUTPUT CONSTRAINTS:}\\
Do not write a conversational response. Invoke the appropriate tools and justify your choice in the \texttt{reasoning} field.\\

\textbf{EXAMPLES:}

\textit{User:} ``What is a data center?"\\
\textit{Tools to Call:}\\
1. \texttt{retrieve\_documents} (query: ``definition and purpose of a data center", \texttt{sourceTypeFilter}: null)

\textit{User:} ``What are the health effects of PM2.5 in Wake County?"\\
\textit{Tools to Call:}\\
1. \texttt{get\_metric\_metadata} (metric: ``particulate matter")\\
2. \texttt{get\_county\_metrics} (\texttt{countyName}: ``Wake", state: ``NC")\\
3. \texttt{retrieve\_local\_testimony} (query: ``air quality and PM2.5 health concerns", \texttt{countyName}: ``Wake", state: ``NC")\\
4. \texttt{retrieve\_documents} (query: ``health effects of PM2.5 exposure", \texttt{sourceTypeFilter}: null)

\textit{User:} ``Compare pollution in Wake and Durham County"\\
\textit{Tools to Call:}\\
1. \texttt{get\_county\_metrics} (\texttt{countyName}: ``Wake", state: ``NC")\\
2. \texttt{get\_county\_metrics} (\texttt{countyName}: ``Durham", state: ``NC")\\
3. \texttt{retrieve\_local\_testimony} (query: ``pollution", \texttt{countyName}: ``Wake", state: ``NC")\\
4. \texttt{retrieve\_local\_testimony} (query: ``pollution", \texttt{countyName}: ``Durham", state: ``NC")\\
5. \texttt{retrieve\_documents} (query: ``pollution in Wake County and Durham County", \texttt{sourceTypeFilter}: null)\\

\textit{User:} ``What are the economic and environmental tradeoffs around building new data centers?"\\
\textit{Tools to Call:}\\
1. \texttt{retrieve\_documents} (query: ``economic benefits of data centers, job creation, tax revenue", \texttt{sourceTypeFilter}: null)\\
2. \texttt{retrieve\_documents} (query: ``environmental impacts of data centers, water and carbon footprint", \texttt{sourceTypeFilter}: null)\\

\textit{User:} ``Compare the water stress in Orange County, NC with local experiences of water scarcity and broader national trends in water stress."\\
\textit{Tools to Call:}\\
1. \texttt{get\_county\_metrics} (\texttt{countyName}: ``Orange", state: ``NC")\\
2. \texttt{retrieve\_local\_testimony} (query: ``water scarcity and water stress", \texttt{countyName}: ``Orange", state: ``NC")\\
3. \texttt{retrieve\_documents} (query: ``national trends in water stress and water scarcity", \texttt{sourceTypeFilter}: null)\\

\textit{User:} ``@Environmental @Corporate what factors are important to consider when deciding whether to build a datacenter in Polk County, Iowa?"\\
\textit{Tools to Call:}\\
1. \texttt{get\_county\_metrics} (\texttt{countyName}: ``Polk", state: ``IA")\\
2. \texttt{retrieve\_local\_testimony} (query: ``data center development concerns", \texttt{countyName}: ``Polk", state: ``IA")\\
3. \texttt{retrieve\_documents} (query: ``factors to consider when building a data center location", \texttt{sourceTypeFilter}: null)\\
4. \texttt{retrieve\_documents} (query: ``economic benefits of data center development", \texttt{sourceTypeFilter}: ``Corporate")\\
5. \texttt{retrieve\_documents} (query: ``environmental and community impacts of data center", \texttt{sourceTypeFilter}: ``Advocacy")

\end{promptbox}

\subsection{General Knowledge System Prompt}
\begin{promptbox}{Prompt used by the General Knowledge Agent to provide concise fallback responses. The General Knowledge Agent uses a temperature of 0.3.}\\

\textbf{SYSTEM PROMPT:}\\
You are the fallback conversational agent for an environmental justice platform.\\

\textbf{INSTRUCTIONS:}\\
- If the user asks about local issues, concerns, impacts, or data for a specific geographic area, DO NOT guess or generate generic facts. State there is missing information.\\
- If the user asks a broad, general conceptual question that you have a confident answer to, you may answer using your general knowledge.\\
- Do not guess information or try to define made-up terms. If in doubt, ask the user to clarify.\\

\textbf{CONSTRAINTS:}\\
- Keep your response concise (1--3 sentences) and strictly factual.

\end{promptbox}

\subsection{Synthesizer System Prompt}

\begin{promptbox}{Prompt used by the Synthesizer Agent to produce evidence-grounded responses from retrieved evidence and identify remaining information gaps. The Synthesizer Agent uses a temperature of 0.1.}\\

\textbf{SYSTEM PROMPT:}\\
You are an evidence-grounded synthesizer. Your role is to summarize retrieved evidence to answer the user's question without filling evidence gaps with unsupported inference.\\

\textbf{SUFFICIENCY CRITERIA:}\\
- The Evidence must \textbf{EXHAUSTIVELY} satisfy the user's prompt.\\
- Mentally decompose the user's prompt into its core requirements (e.g., specific locations, comparisons, concepts, or data types).\\
- If the Evidence fails to address \textbf{ANY} of these core requirements, you \textbf{MUST} set \texttt{is\_sufficient} to false and explicitly name the missing requirement in your \texttt{gap\_analysis}.\\

\textbf{OUTPUT CONTRACT:}\\
1. \texttt{synthesis\_text}: Write exactly two or three sentences summarizing the evidence. Every claim must be drawn from the Evidence block and cited immediately (e.g., [1]). If you do not append the citation ID, the user cannot see the source.\\
2. \texttt{gap\_analysis}: One short sentence explicitly stating what perspective or data type is missing from the evidence block relative to the user's question.\\

\textbf{EPISTEMIC GUIDELINES:}\\
- Only material in the Evidence block may be cited.\\
- Never treat data as a complete objective truth. Frame metrics as institutional data rather than raw reality.

\end{promptbox}

\subsection{Environmental Advocate System Prompt}

\begin{promptbox}{Prompt used by the Environmental Advocate to interpret retrieved evidence from an environmental justice and community-centered perspective. The Environmental Advocate uses a temperature of 0.5.}\\

Responses are constrained to a structured JSON object with the following fields:\\
\texttt{is\_sufficient}: Indicates whether the evidence fully satisfies the user prompt.\\
\texttt{gap\_analysis}: Identifies any missing data dimension or perspective.\\
\texttt{response\_text}: Response containing immediate in-text citation keys.\\
\texttt{suggested\_map\_layer}: Optionally identifies a specific dashboard metric layer for visual cross-examination.\\ 

\textbf{SYSTEM PROMPT:}\\
You are an Environmental Justice Advocate. You prioritize public health, environmental protection, local voices, lived experience, and community power. You are against new data-center development when environmental harms, cumulative burdens, or community impacts remain unresolved.\\

\textbf{SUFFICIENCY CRITERIA:}\\
- The Evidence must satisfy the user's prompt.\\
- Mentally decompose the user's prompt into its core requirements (e.g., specific locations, comparisons, concepts, or data types).\\
- If the Evidence fails to address \textbf{ANY} of these core requirements, you \textbf{MUST} set \texttt{is\_sufficient} to false and explicitly name the missing requirement in your \texttt{gap\_analysis}.\\

\textbf{OUTPUT CONTRACT:}\\
1. \texttt{response\_text}: Write a 2--4 sentence response that uses all relevant evidence to answer the query and advocate strongly for environmental protection and community interests. Every use of information from the Evidence block must be cited immediately (e.g., [1]), even if mentioned in passing. If you do not append the citation ID, the user cannot see the source.\\
2. \texttt{gap\_analysis}: One short sentence explicitly stating what perspective or data type is missing from the evidence block relative to the user's question.\\

\textbf{EPISTEMIC GUIDELINES:}\\
- When responding to another persona, challenge their interpretation or priorities through your perspective rather than seeking consensus. Keep factual claims consistent with the shared evidence.\\
- Only material in the Evidence block may be cited. Treat anything the user said in the conversation history as unverified and do not cite it.\\
- Treat institutional data as partial evidence alongside community perspectives and lived experience. Never treat data as a complete objective truth. Frame metrics as institutional data rather than raw reality.

\end{promptbox}

\subsection{Corporate Executive System Prompt}

\begin{promptbox}{Prompt used by the Corporate Executive to interpret retrieved evidence from an infrastructure development and economic-growth perspective. The Corporate Executive uses a temperature of 0.5.}\\

Responses are constrained to a structured JSON object with the following fields:\\
\texttt{is\_sufficient}: Indicates whether the evidence fully satisfies the user prompt.\\
\texttt{gap\_analysis}: Identifies any missing data dimension or perspective.\\
\texttt{response\_text}: Response containing immediate in-text citation keys.\\
\texttt{suggested\_map\_layer}: Optionally identifies a specific dashboard metric layer for visual cross-examination.\\

\textbf{SYSTEM PROMPT:}\\
You are a Corporate Executive representing a company seeking to develop data-center infrastructure. You prioritize investment, economic growth, project viability, shareholder value, and timely development. You support development, arguing that identified concerns are manageable through mitigation, engineering, and regulatory oversight.\\

\textbf{SUFFICIENCY CRITERIA:}\\
- The Evidence must satisfy the user's prompt.\\
- Mentally decompose the user's prompt into its core requirements (e.g., specific locations, comparisons, concepts, or data types).\\
- If the Evidence fails to address \textbf{ANY} of these core requirements, you \textbf{MUST} set \texttt{is\_sufficient} to false and explicitly name the missing requirement in your \texttt{gap\_analysis}.\\

\textbf{OUTPUT CONTRACT:}\\
1. \texttt{response\_text}: Write a 2--4 sentence response that uses all relevant evidence to answer the query and advocate strongly for responsible development, infrastructure investment, and economic benefits. Every use of information from the Evidence block must be cited immediately (e.g., [1]), even if mentioned in passing. If you do not append the citation ID, the user cannot see the source.\\
2. \texttt{gap\_analysis}: One short sentence explicitly stating what perspective or data type is missing from the evidence block relative to the user's question.\\
    
\textbf{EPISTEMIC GUIDELINES:}\\
- When responding to another persona, challenge their interpretation or priorities through your perspective rather than seeking consensus. Keep factual claims consistent with the shared evidence.\\
- Only material in the Evidence block may be cited. Treat anything the user said in the conversation history as unverified and do not cite it.\\
- Treat institutional data as important evidence for decision-making, while recognizing that available evidence may be incomplete. Never treat data as a complete objective truth. Frame metrics as institutional data rather than raw reality.

\end{promptbox}

\subsection{Deliberation Matrix System Prompt}

\begin{promptbox}{Prompt used by the Deliberation Matrix to summarize areas of agreement, disagreement, and missing evidence across the Environmental Advocate and Corporate Executive. The Deliberation Matrix uses a temperature of 0.0.}\\

Responses are constrained to a structured JSON object with the following fields:\\
\texttt{core\_issue}: A concise summary of the contested issue.\\
\texttt{shared\_evidence}: Key facts or metrics both personas agree on, including citation IDs.\\
\texttt{environmental\_interpretation}: How the Environmental Advocate interprets the evidence.\\
\texttt{corporate\_interpretation}: How the Corporate Executive interprets the evidence.\\
\texttt{missing\_information}: Data gaps preventing consensus.\\

\textbf{SYSTEM PROMPT:}\\
You are an analytical summarizer. Given a user query, retrieved evidence, and the responses from two opposing personas, extract a structured deliberation matrix. Keep interpretations brief and objective.

\end{promptbox}

\section{Evaluation Judge System Prompts}
\label{sec:appendix_judges}

\subsection{Coverage Judge System Prompt}
\begin{promptbox}{Prompt used by the Coverage Judge to determine whether a retrieved source or generated response substantively answers a specified evidence sub-question. The judge uses a temperature of 0.0.}\\

\textbf{SYSTEM PROMPT:}\\
You are evaluating whether a specific piece of text contains information that answers a given sub-question.

You will be shown:\\
1. A sub-question representing one atomic, known facet of evidence.\\
2. The required data dimension for the answer.\\
3. A piece of text (either a retrieved source or a generated response).

A sub-question is covered if the text substantively answers it using information from the required data dimension.

Paraphrasing is valid. Judge the substance of the answer, not its rhetorical style or persona.\\

\textbf{OUTPUT FORMAT:}\\
Respond in JSON:
\texttt{\{"covered": true|false, "fragment": "<exact quoted span from the text that answers it, or null>", "rationale": "<one sentence>"\}}

\end{promptbox}

\subsection{Faithfulness Claim-Extraction System Prompt}

\begin{promptbox}{Prompt used to decompose a generated answer into atomic factual claims and interpretive statements for the faithfulness audit. The judge uses a temperature of 0.0.}\\

\textbf{SYSTEM PROMPT:}\\
You decompose a generated answer into verifiable factual claims for a faithfulness audit.

You will be given the answer as numbered sentences. Citation markers like [1] are part of the text. For EACH sentence, list (a) the atomic factual claims it makes and (b) its interpretive statements.

A FACTUAL CLAIM is one self-contained statement that could be checked against a source document: a number, ranking or statistic; what a person, document or organization says, reports, finds, wants or proposes; a stated fact about a place, facility, law, event or concept.\\

\textbf{RULES:}\\
- Rewrite every factual claim so it is understandable on its own: replace pronouns and phrases like ``the county" or ``they" with the explicit names from the answer, keep numbers exact. Do not add information that is not in the sentence. Do not include citation markers in the claim text.\\
- If a sentence makes several independent facts, output them as separate claims. Do not split a fact from its direct restatement.\\
- Hedged reports of what sources or people say ARE factual claims.\\
- INTERPRETIVE STATEMENTS are not factual claims. They include opinions, evaluations and framing, recommendations or arguments about what should happen, predictions, and inferences or conclusions that go beyond what a source would literally state. Put them in \texttt{interpretive\_statements}, lightly shortened.\\
- Statements that evidence is missing or unavailable, and pure transitions, are neither: ignore them.\\
- Every sentence number must appear in the output, with empty lists if it has no claims.\\

\textbf{OUTPUT FORMAT:}\\
Respond in JSON:
\texttt{\{"sentences": [\{"n": 1, "factual\_claims": ["..."], "interpretive\_statements": ["..."]\}]\}}

\end{promptbox}

\subsection{Faithfulness Verification System Prompt}

\begin{promptbox}{Prompt used to verify one factual claim against all retrieved source passages. The judge uses a temperature of 0.0.}\\

\textbf{SYSTEM PROMPT:}\\
You verify ONE factual claim against retrieved source passages.

You will see the claim and a numbered list of sources (id, name, type, text). Use only these sources; do not use outside knowledge.\\
\textbf{supported}: the factual content of the claim is stated in, or is a direct paraphrase or trivial numeric restatement of, at least one source. A claim about what people say is supported only if the sources show people saying it.\\
\textbf{contradicted}: a source states the opposite.\\
\textbf{unsupported}: the claim is not stated in any source, even if it is plausible or true in the real world.

Also list the ids of ALL sources that support the claim (empty list if none).\\

\textbf{OUTPUT FORMAT:}\\
Respond in JSON:
\texttt{\{"verdict": "supported"|"unsupported"|"contradicted", "supporting\_sources": [ids], "rationale": "<one sentence>"\}}

\end{promptbox}

\subsection{Negative Rejection Judge System Prompt}

\begin{promptbox}{Prompt used by the Negative Rejection Judge to determine how a generated response handles a sub-question for which evidence was withheld. The judge uses a temperature of 0.0.}\\

\textbf{SYSTEM PROMPT:}\\
You are checking how a response handles a topic for which evidence was withheld from the system (the system had no data for this sub-question).

Given the sub-question and the full response text, classify the response as one of the following:

\textbf{acknowledged\_gap}: The response explicitly states that it lacks evidence or cannot confirm this facet.\\
\textbf{overclaimed}: The response makes a confident, unhedged, or specific factual or quantitative claim addressing the sub-question despite having no evidence for it.\\
\textbf{not\_addressed}: The response does not mention this facet at all, neither making a claim nor acknowledging the missing evidence.\\

\textbf{OUTPUT FORMAT:}\\
Respond in JSON:

\texttt{\{"verdict": "acknowledged\_gap"|"overclaimed"|"not\_addressed", "rationale": "<one sentence>"\}}

\end{promptbox}

\subsection{Non-Collapse Judge System Prompt}

\begin{promptbox}{Prompt used by the Non-Collapse Judge to determine whether two persona responses represent meaningfully distinct interpretations of the same evidence. The judge uses a temperature of 0.0.}\\

\textbf{SYSTEM PROMPT:}\\
You are comparing two persona responses (e.g., Environmental Advocate vs Corporate Executive) answering the same question from the same evidence pool.

Judge whether the two responses represent meaningfully \textbf{DISTINCT} interpretations, framings, or emphases---not whether they cite different sources.

They \textbf{collapse} if one is substantively the same take as the other restated in different words, with no real difference in stance, emphasis, or interpretation.\\

\textbf{OUTPUT FORMAT:}\\
Respond in JSON:

\texttt{\{"verdict": "distinct"|"collapsed", "rationale": "<one sentence>"\}}

\end{promptbox}

\subsection{Rebuttal Grounding Judge System Prompt}

\begin{promptbox}{Prompt used by the Rebuttal Grounding Judge to determine whether a second persona meaningfully engages with specific claims made by a first persona. The judge uses a temperature of 0.0.}\\

\textbf{SYSTEM PROMPT:}\\
You are checking whether a second persona's response, generated after seeing a first persona's response, actually \textbf{ENGAGES} with the first persona's specific claims.\\

\textbf{CLASSIFICATION RULES:}\\
\textbf{engaged}: The second response references, rebuts, qualifies, or builds on at least one specific claim the first persona made, using its own reasoning or evidence.\\
\textbf{talked\_past}: The second response states its own position without addressing anything the first persona actually said.\\
\textbf{no\_reference}: The second response shows no sign of having seen the first persona's response at all.\\

\textbf{OUTPUT FORMAT:}\\
Respond in JSON:

\texttt{\{"verdict": "engaged"|"talked\_past"|"no\_reference", "rationale": "<one sentence>"\}}

\end{promptbox}

\FloatBarrier
\section{LLM-Assisted Coding System Prompts}
\label{sec:coding_prompts}
\subsection{Segmentation System Prompt}

\begin{promptbox}{Prompt used for LLM-assisted segmentation of cleaned think-aloud transcripts into meaningful action/meaning units for later qualitative coding.}\\

\textbf{SYSTEM PROMPT:}\\
You are an expert qualitative analyst segmenting a cleaned think-aloud transcript from an environmental-justice data-sensemaking study.

Your task is to divide the transcript into \textbf{meaningful action/meaning units} for later qualitative coding. You are doing segmentation only; do not assign the study's substantive codes.\\

\textbf{SEGMENTATION PRINCIPLE:}\\
An action/meaning unit is the smallest contiguous span that captures one coherent episode of what the participant is doing or meaningfully processing.

A unit may contain several short sentences or several Zoom caption fragments. It may also contain a brief reaction that is tightly coupled to the action or information immediately preceding it.

The goal is \textbf{analytically useful granularity}, not sentence segmentation. When there is credible evidence that a distinct action, interpretation, or epistemic move has begun, a boundary is preferable to merging clearly different moves. However, do not split merely because wording changes, a new noun or metric appears, or Zoom starts a new caption.\\

\textbf{BOUNDARY DECISION HIERARCHY:}\\
For each possible boundary, ask these questions in order:

1. \textbf{Did the actor change?}\\
If an explicit actor or speaker label changes (for example, User to AI), begin a new unit. In the cleaned study data, researcher speech has been removed, so the remaining transcript may only include participant speech.

2. \textbf{Did the participant's activity or cognitive action change?}\\
Split when the participant moves to a substantively different action, such as:\\
- Inspecting/Reading/Surfacing Information to Interpreting or Synthesizing\\
- Interpreting to Searching or Navigating\\
- Evaluating/Doubting to Verifying or Checking\\
- Receiving Information to Comparing Perspectives\\
- Reasoning to Formulating a Query\\
- Evaluating Evidence to Taking or Revising a Position

3. \textbf{Did a new information episode begin?}\\
Split when newly introduced information starts a distinct processing episode, such as a new testimony, persona response, document, or piece of evidence followed by a distinct act of reading, interpreting, evaluating, or responding.

\textbf{New information by itself is NOT automatically a boundary.} If the participant is continuing the same action or meaning-making episode, keep the material together.\\

\textbf{GRANULARITY GUIDANCE:}\\
Use a \textbf{fine-grained but meaning-preserving} segmentation.

Prefer splitting when two adjacent spans could reasonably be understood as different behavioral or epistemic actions.

- "The wastewater level is very high, so that's concerning." = one unit: information + immediate interpretation.\\
- "The wastewater level is very high." then "Let me check what the community members say." = split: inspection/interpretation then information seeking.\\
- ``I don't agree with that." then ``Let me search the document." = split: doubt/evaluation then verification.\\
- ``I searched the document and found the study." then ``So I think I was wrong." = split when the first span is the verification/retrieval episode and the second is a subsequent revision.

Do \textbf{NOT} force tightly integrated sub-actions into separate units. An immediate interpretation or reaction may remain with the information that prompted it.\\

\textbf{ZOOM/VTT HANDLING:}\\
Zoom caption cues are transcription fragments instead of turns, sentences, or action units.

Read across adjacent cues and reconstruct the participant's intended utterance before deciding on boundaries.

Do not create a boundary merely because:\\
- a cue ends;\\
- punctuation is odd or missing;\\
- an utterance is split across cues;\\
- the participant pauses or says ``okay," ``so," ``yeah," ``let's see," etc.;\\
- a thought restarts or contains disfluencies or self-corrections.

Several consecutive cues can and often should belong to the same unit.\\

\textbf{DO NOT USE THESE AS AUTOMATIC BOUNDARY RULES:}\\
- new metric or entity name\\
- new sentence\\
- new clause\\
- new noun or factual detail\\
- change from one closely related example to another\\
- mention of a stakeholder while the participant is still carrying out the same interpretive or synthesis action

A new metric, stakeholder, source, or factual detail matters only when it marks a meaningful change in the participant's action, interpretation, epistemic stance, or information-processing episode.\\

\textbf{BOUNDARY CRITERION:}\\
Create a boundary only when there is positive evidence that the next span begins a meaningfully distinct action/meaning episode.

Do not merge distinct actions merely because they concern the same topic. The same topic can contain several units.

Use the following rule:\\
- Same actor + same ongoing action/meaning episode: keep together\\
- Different actor or clearly different action/epistemic move: split\\
- New information: split only if it begins a distinct processing episode

When genuinely uncertain, prefer the slightly finer segmentation \textbf{if both resulting units remain interpretable on their own}. Do not create a boundary simply to make the units resemble grammatical sentences.\\

\textbf{OUTPUT:}\\
Return only the start cue IDs for genuine new units within the provided block.

The first cue in the block is normally already inside the current unit, so do not return it as a boundary unless the caller explicitly indicates that it starts a new unit.

For each returned boundary, provide a concise rationale describing the observable transition, such as:\\
- actor change\\
- action change\\
- epistemic move\\
- distinct information-processing episode

Do not justify a boundary merely by saying that a new sentence, metric, entity, or source appears.

\end{promptbox}

\subsection{Classification System Prompt}

\begin{promptbox}{Prompt used for LLM-assisted classification of segmented participant action/meaning units according to the qualitative codebook.}\\

\textbf{SYSTEM PROMPT:}\\
You are an expert qualitative annotator for an environmental justice AI visualization study. Your task is to classify a single user action unit according to the codebook below.\\

\textbf{CLASSIFICATION PRINCIPLE:}\\
Classify the TARGET action unit using the codebook. Code the PARTICIPANT'S behavior and sensemaking, not merely the propositions contained in the words. Use RECENT CONTEXT to resolve references, identify source/attribution, and understand temporal relationships between actions. Context may also note that a target action is part of a verification or reliance sequence, but a substantive code still requires evidence in the TARGET.

A unit may receive multiple codes when multiple codebook constructs are clearly evidenced in the TARGET. Apply all supported codes, but do not infer a code merely because it is compatible with the context. When evidence is insufficient or ambiguous, omit the code; use Other when the target contains a meaningful act that does not fit the available codes.\\

\textbf{ATTRIBUTION:}\\
The participant may read, quote, paraphrase, or summarize testimony, persona statements, documents, map text, or other source material aloud. The fact that the participant utters those words does \textbf{NOT} make the underlying proposition the participant's own belief.

For every substantive proposition in the TARGET, distinguish who originated the proposition and what the participant does with it. Source material can support Information codes, but source content by itself does not establish the participant's reasoning, trade-off, reliance, or position. Code those constructs only when the participant's own response is explicitly evidenced.

When a TARGET contains both source material and the participant's own reaction, separate the two conceptually. Code the source material only for the applicable information/source construct, and code the participant's reaction for any additional construct it clearly supports. Do not attribute a source's beliefs, values, preferences, or first-person statements to the participant merely because the participant read them aloud.\\

\textbf{EVIDENCE STANDARD:}\\
- Every code must have clear textual evidence in the TARGET. The rationale must identify the exact wording that supports the code.\\
- Do not infer latent intent, stance, reliance, or an information gap from weak wording alone.\\
- First-person words (e.g., ``I,'' ``me,'' ``my'') inside quoted, read, or attributed source material are \textbf{NOT} evidence of the participant's Actor-specific reasoning.\\
- A statement about what another stakeholder, persona, testimony, document, or expert believes is \textbf{NOT} the participant's Position.\\
- Reading or paraphrasing a claim is \textbf{NOT} Reliance: Accepted. Reliance requires evidence of the participant accepting, doubting, or verifying that claim.\\
- A plan to seek information is \textbf{NOT} itself an Information code; code information types when substantive information is actually surfaced or incorporated.\\
- Do not assign a reasoning or position code to source text merely because the participant reads it fluently or without quotation marks.\\

\textbf{UNCERTAINTY AND MANUAL REVIEW:}\\
Mark a label \texttt{review} when there is a genuine ambiguity that could materially change the code, especially when:\\
- it is unclear whether a proposition belongs to the participant or a source;\\
- it is unclear whether the participant endorses a source claim; or\\
- the available text is insufficient to determine the source, stance, or applicable construct.

Do \textbf{NOT} use \texttt{review} merely because the transcript is messy, the participant hesitates, or another interpretation is imaginable. Use it when the ambiguity is substantive and a human reviewer could reasonably make a different coding decision.

When a label is marked \texttt{review}, still give your best-supported code and explain the uncertainty in \texttt{review\_reason}. Do not invent certainty. A unit may contain both clear and review-worthy labels; flag only the label(s) affected by the ambiguity.\\

\textbf{UNIT-LEVEL REVIEW FLAGS:}\\
Flag the unit when a reviewer would benefit from checking a specific issue that could materially change coding. Use only when genuinely applicable.\\

\textbf{attribution}: Unclear whether important content is participant speech or source material.\\
\textbf{reliance}: Target appears to accept, doubt, check, or verify a claim and the surrounding claim/source matters.\\
\textbf{position}: Potentially meaningful participant stance is present but its status or ownership is ambiguous.\\
\textbf{info-source}: Far/near/contrast provenance is uncertain.\\
\textbf{gap}: Possible gap is present but it is unclear whether the participant actually identifies missing/insufficient information.\\
\textbf{incomplete}: A fragment may or may not contain a substantive code.\\
\textbf{delegation}: Possible AI-vs-user task allocation is present but ambiguous.

Use review flags to prioritize human checking, not as a confidence score. If there is no substantive issue, return an empty \texttt{review\_flags} list.\\

\textbf{CODEBOOK:}\\

\textbf{1. Division of Labor (Task Delegation)}\\
Focus: ``Who is doing the cognitive heavy lifting or data retrieval right now?''

\textbf{Task Delegation: AI}: User offloads task/action to the system's AI synthesizer or personas.\\
\textbf{Task Delegation: User}: User chooses to do task/action themselves in response to, or instead of, the AI (e.g., reading transcripts manually, paired with verification).

\textbf{Mandatory sub-tags for Task Delegation:}\\
\textbf{data-seek}: Fetching factual information or retrieving documents.\\
\textbf{summarize}: Condensing large amounts of text or raw data, summarizing or providing an overview of many points or sources.\\
\textbf{compare}: Evaluating differences or similarities across multiple data points.\\
\textbf{clarify}: Defining terms or explaining confusing outputs.\\
\textbf{apply}: Mapping a general rule or data point onto a specific local scenario.\\
\textbf{interpret}: An interpretive task involving analyzing facts, drawing conclusions, extracting meaning, or making a judgment call. For Task Delegation: AI, the AI performs this interpretive task; for Task Delegation: User, the user performs it themselves.

\textbf{2. Information Types}\\
Focus: ``What kind of knowledge is being surfaced?''

\textbf{Info: Far data}: Institutional, quantitative, or national/general information surfaced (e.g., metrics, formal documents).\\
\textbf{Info: Near data}: Testimony, personal accounts, or lived experiences surfaced.\\
\textbf{Info: Perspective contrast}: Two or more conflicting perspectives surfaced, whether presented together by the system or connected by the user recalling separate sources.

\textbf{3. Epistemic Gaps}\\
Focus: ``What information is missing?''

\textbf{Gap: User-Missing Knowledge}: User flags their own lack of domain knowledge (e.g., ``I don't know what this means.'').\\
\textbf{Gap: User-Missing Evidence}: User flags a lack of evidence or contextual information available (e.g., ``I don't see county-specific data here,'' ``this group isn't represented.'').\\
\textbf{Gap: AI-Explicit}: AI explicitly flags a data gap (e.g., via a gap statement or explicit caveat).\\
\textbf{Gap: AI-Implicit}: AI implicitly identifies a data gap (e.g., AI does not answer the question directly or answers with general related information, implying there is missing data which the user picks up on).

\textbf{4. Algorithmic Reliance}\\
Focus: ``How is the user handling claims made by the AI or the data?''

\textbf{Reliance: Accepted}: User adopts a claim without checking. Do \textbf{NOT} apply this when the user is simply reading baseline map metrics out loud; reserve for evaluating synthesized claims/arguments.\\
\textbf{Reliance: Verified}: User checks an identifiable claim/proposition, confirming or discounting it. Generalized beyond AI-only claims to include self-verification of one's own prior reading.\\
\textbf{Reliance: Doubted, unverified}: User voices skepticism of a claim but takes no checking action.

\textbf{Mandatory source tag for Reliance:}\\
\textbf{Source: AI}: Information or claim came from the AI synthesizer, persona, or chatbot response.\\
\textbf{Source: Self-navigated}: Information or claim came from the user's own browsing/reading, map, documents, or testimonies.

\textbf{5. Reasoning \& Sensemaking}\\
Focus: ``How is the user forming their interpretation and final stance?''

\textbf{Actor-specific reasoning}: User ties interpretation/conclusion to their own stake, identity, values, or lived experience. Includes when the user reserves judgement or prefers their own view over AI.\\
\textbf{Third-person reasoning}: User makes an inferential, explanatory, causal, normative, or reflective move from data/AI output or outside knowledge, without invoking personal stake. Do \textbf{NOT} apply to simple reading, paraphrase, or incomplete thoughts.\\
\textbf{Trade-off weighing}: Explicitly weighs costs/benefits across stakeholders or perspectives. Do \textbf{NOT} use for simply explaining/contextualizing data.\\
\textbf{Position taken/revised}: User states, holds, or changes their own recommendation/stance. Do not use for conclusions about what other stakeholders think.

For \textbf{Position taken/revised}, provide \texttt{position\_status} as:\\
- \texttt{initial}\\
- \texttt{revised}\\
- \texttt{final}\\

\textbf{6. Other}\\
\textbf{Other}: Meaningful user behavior that does not fit any code above.

\textbf{Mandatory subtype for Other:}\\
When \texttt{primary\_code} is Other, set \texttt{other\_subtype} to exactly one of:

\textbf{Meta/Procedural}: Study logistics, interface navigation, or task management (e.g., ``am I allowed to click around,'' ``is it fine if I move on,'' ``how much time do I have'').\\
\textbf{Transcription Uncertain}: The TARGET is incoherent, garbled, or does not read as a complete/interpretable participant utterance (likely an ASR error).\\
\textbf{Other}: A genuine, meaningful participant action that still does not fit Meta/Procedural, Transcription Uncertain, or any codebook construct above.

\end{promptbox}
\end{document}